\pdfoutput=1
 
\documentclass[preprint,3p,times]{elsarticle}

\usepackage{graphicx}
\usepackage{epsfig}

\usepackage{amssymb}

\usepackage[nodots]{numcompress}

\usepackage{xcolor}

\usepackage{epsfig,lineno,textcomp,subfigure,units}

\newcommand{\pienu}{$\pi^{+} \rightarrow e^{+} \nu_{e}$ }
\newcommand{\pimue}{$\pi^+ \to \mu^{+}\nu_{\mu}$}

\newcommand{\BR}{${\Gamma\big((\pi^{+} \rightarrow e^{+} \nu_{e}) + (\pi^{+} \rightarrow e^{+} \nu_{e}\gamma)\big)}/{\Gamma\big((\pi^{+} \rightarrow \mu^{+} \nu_{\mu})+(\pi^{+} \rightarrow \mu^{+} \nu_{\mu}\gamma)\big)}$ }

\newcommand{\BRs}{$R^{e/\mu}$}

\journal{Nuclear Inst. and Methods in Physics Research, A}
\begin{document}

\begin{frontmatter}


\title{Detector for measuring the $\pi^+\to e^+\nu_e$ branching fraction}
\author[label0]{A. A. Aguilar-Arevalo}
\author[label1]{M. Aoki}
\author[label2]{M. Blecher}
\author[label4]{D. vom Bruch\fnref{fn1}}
\author[label4]{D. Bryman}
\author[label9]{J. Comfort}
\author[label4]{S. Cuen-Rochin}
\author[label3]{ L. Doria\corref{cor1}}
\ead{luca@triumf.ca}
\author[label3]{ P. Gumplinger}
\author[label6]{ A. Hussein}
\author[label7]{ Y. Igarashi}
\author[label1]{ N. Ito}
\author[label1]{ S. Ito}
\author[label8]{ S. H. Kettell}
\author[label3]{ L. Kurchaninov}
\author[label8]{L. Littenberg}
\author[label4]{ C. Malbrunot\corref{cor1}\fnref{fn2}}
\ead{chloe.m@cern.ch}
\author[label3]{R.E. Mischke}  
\author[label1]{A. Muroi}
\author[label3]{T. Numao}
\author[label3]{G. Sheffer}
\author[label3]{A. Sher\corref{cor2}}
\ead{sher@triumf.ca}
\author[label4]{T. Sullivan}
\author[label7]{K. Tauchi}
\author[label3]{D. Vavilov}
\author[label1]{K. Yamada \fnref{fn3}}
\author[label1]{M. Yoshida \fnref{fn4}}

\cortext[cor1]{Corresponding author.}
\cortext[cor2]{Primary corresponding author.}

\fntext[fn1]{Present address: Physikalisches Institut, Universität Heidelberg, Heidelberg, Germany.}
\fntext[fn2]{Present address: CERN, 1211 Geneva 21, Switzerland and Stefan-Meyer-Institut f\"ur subatomare Physik, Austrian Academy of Sciences, Boltzmanngasse 3, A-1090 Vienna, Austria.}
\fntext[fn3]{Present address: Nagoya Koyo Senior High School, Japan.}
\fntext[fn4]{Present address: KEK, 1-1 Oho, Tsukuba-shi, Ibaragi, Japan.}




\address[label0]{Instituto de Ciencias Nucleares, Universidad Nacional Aut\'onoma de Mexico, D.F. 04510 M\'exico}
\address[label1]{Graduate School of Science, Osaka University, Toyonaka, Osaka, 560-0043, Japan}
\address[label2]{Virginia Polytechnic Institute and State University, Blacksburg, VA, 24061, USA}
\address[label3]{TRIUMF, Vancouver, B.C., V6T 2A3, Canada}
\address[label4]{Department of Physics and Astronomy, University of British Columbia, Vancouver, B.C., V6T 1Z1, Canada}
\address[label6]{University of Northern British Columbia, Prince George, B.C., V2N 4Z9, Canada}
\address[label7]{KEK, Tsukuba-shi, Ibaragi, Japan}
\address[label8]{Brookhaven National Laboratory, Upton, NY, 11973-5000, USA}
\address[label9]{Arizona State University, Tempe, AZ  85287-1504, USA}

\begin{abstract}
The PIENU experiment at TRIUMF is aimed at a measurement of the branching ratio\\ \BRs=\BR with precision $<$0.1\%.  
Incident pions, delivered at the rate of 60 kHz with momentum 75 MeV/c, were degraded and stopped in a 
plastic scintillator target.
 Pions and their decay product positrons were detected with plastic scintillators and tracked with multiwire proportional
 chambers  and silicon strip detectors. The energies of the positrons were  measured in a spectrometer consisting of
 a large NaI(T$\ell$) crystal surrounded by an array of pure CsI crystals.
This paper provides a description of the PIENU experimental apparatus and its performance in pursuit of \BRs.

\end{abstract}


\end{frontmatter}

\section{Introduction}
\label{sec:intro}

Measurement of the branching ratio \BRs 
\begin{displaymath}
R^{e/\mu}=\frac{\Gamma\big((\pi^{+} \rightarrow e^{+} \nu_{e}) + (\pi^{+} \rightarrow e^{+} \nu_{e}\gamma)\big)}{\Gamma\big((\pi^{+} \rightarrow \mu^{+} \nu_{\mu})+(\pi^{+} \rightarrow \mu^{+} \nu_{\mu}\gamma)\big)}
\end{displaymath}
compared to its precise Standard Model (SM) prediction
provides a
stringent test of the hypothesis of  $e-\mu$ universality in weak interactions. 
The most recent experimental measurements of the branching ratio are
\begin{displaymath}
 R^{e/\mu}_{\textrm{\tiny{TRIUMF}}}=(1.2265 \pm
0.0034(stat) \pm 0.0044(syst)) \times 10^{-4} \textrm{~\cite{PRL92,PRD94}},
\end{displaymath}  and
\begin{displaymath}
\noindent{
R^{e/\mu}_{\textrm{\tiny{PSI}}}=(1.2346 \pm 0.0035(stat) \pm 0.0036(syst)) \times 10^{-4} \textrm{~\cite{PSI93}}.}
\end{displaymath}
These measurements are in agreement with the prediction:
\begin{displaymath}
 R^{e/\mu}_{\textrm{\tiny{SM}}}=1.2352(1)\times 10^{-4} \textrm{~\cite{th1,th2}}. 
\end{displaymath}
The order of magnitude gap in the precision of the experimental measurements and the theoretical prediction 
provides an opportunity to test the SM and to search for physics beyond it.
Because of helicity-suppression of $\pi^+ \to e^+\nu_{e}$ decay in the SM, helicity-unsuppressed contributions due to
 pseudoscalar or scalar couplings not present in the SM could result in a deviation of \BRs $ $ from the
 SM prediction. A measurement of  \BRs $ $ with $<$0.1\% precision can probe new physics
 at mass scales 
up to
1000 TeV for pseudoscalar interactions ~\cite{Bryman}.

The PIENU
experimental
 technique 
is based on observing
 positrons from decays of  pions at rest in an active plastic scintillator target. Muons from the dominant \pimue $ $ decay, which have kinetic energy
of 4.1 MeV and range of about 1~mm, were stopped in the target.
By measuring the time and energy of positrons from the decays \pienu and 
 $\mu^+\rightarrow e^+\nu_e \bar{\nu}_{\mu}$  following the decay $\pi^+\rightarrow\mu^+\nu_{\mu}$ ($\pi^+\to\mu^+\to e^+$ decay chain),
 $R^{e/\mu}$ can be obtained after applying several corrections. 
Since the positrons from the \pienu decay are monochromatic ($T_{\pi \to e \nu}$ = 69.3 MeV), and positrons
 from the $\pi^+\to\mu^+\to e^+$ decay chain have a three-body decay energy spectrum with highest positron kinetic energy cutoff at 52.3 MeV, 
the measured energy is used to separate the two decay modes.
 The raw branching ratio is extracted by performing a simultaneous fit of the time spectra of high energy ($T_e >$52 MeV) and low energy ($T_e <$52 MeV) samples 
to determine the yields of \pienu and  $\pi^+\to\mu^+\to e^+$ decay chain events while taking backgrounds into account.  Normalization factors, such as the solid angle of positron detection, cancel to first order, 
and the measured decay yields must be corrected for small energy-dependent
 effects, such as those from multiple Coulomb scattering, Bhabha scattering, 
and positron annihilation. For the final branching ratio determination additional corrections are applied including
 one which accounts for those \pienu decays excluded from the high energy time spectrum due to electromagnetic shower leakage and other effects. The latter correction and its systematic uncertainty were empirically determined from data
 and  measurements of the response of the calorimetry system. 

The PIENU detector had a solid angle acceptance\footnote{Acceptance was defined by two plastic scintillator counters located downstream of the pion stopping target.} for positrons emitted from the stopping target of
 25\%, which is eight times higher than the previous TRIUMF experiment \cite{PRD94}.
In addition to measuring \BRs, because of the increased acceptance and better energy resolution, 
 the PIENU 
experiment was also able to perform a more sensitive search for heavy neutrinos in the $\pi^+\to e^+\nu_{H}$ decay \cite{mnu}.

\section{Experimental overview}

\begin{figure}[!ht]
\begin{center}
\includegraphics[trim=0cm 0cm 0cm 0cm, clip=true,angle=-90,width=0.8\columnwidth]{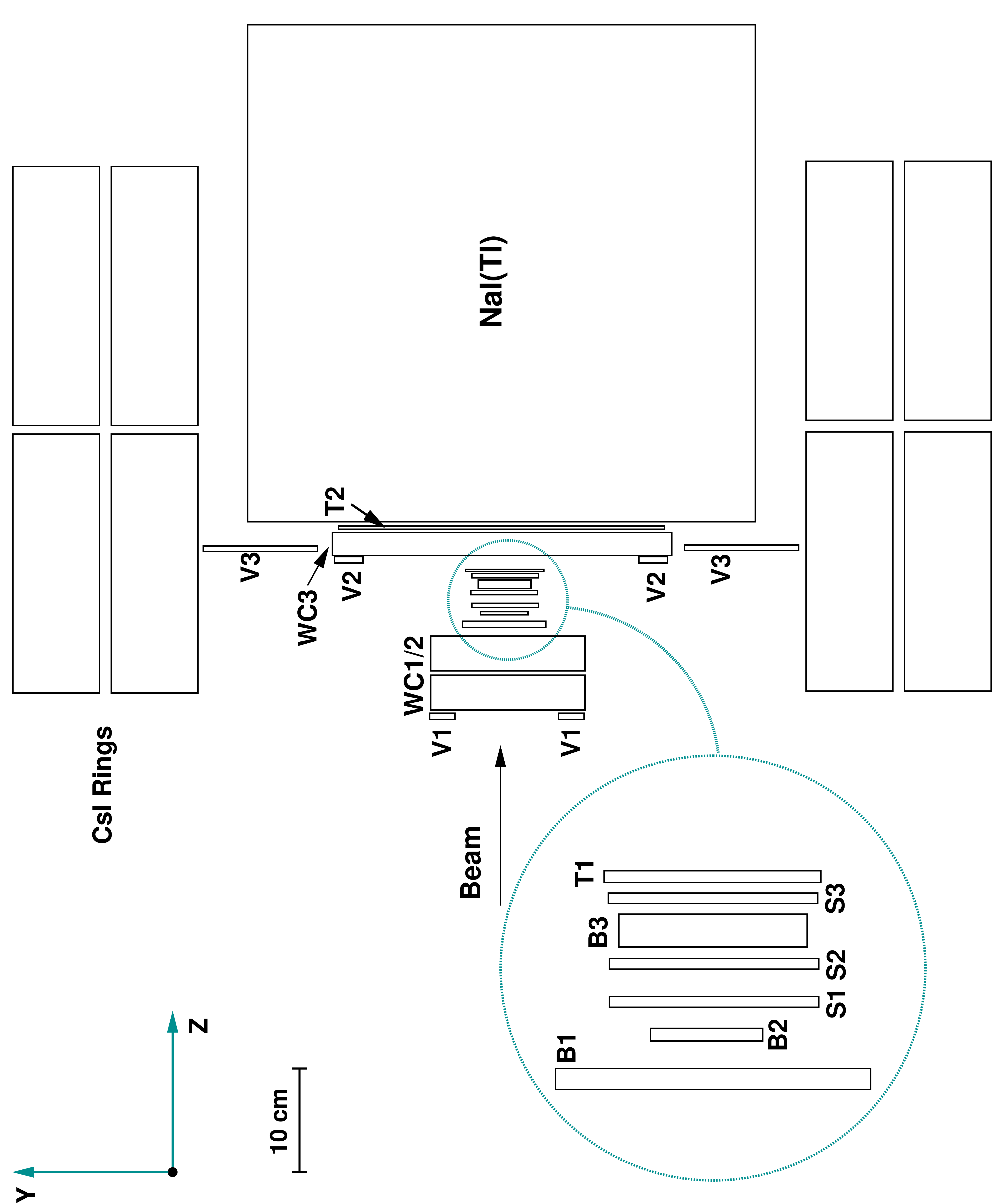}
\caption[Schematic diagram of the experimental apparatus.]{Schematic diagram of the experimental apparatus. See text.}
\label{detview}
\end{center}
\end{figure}

The TRIUMF cyclotron delivered a 500~MeV proton beam with an intensity of
 120~$\mu$A to a 12~mm thick beryllium pion production target.
A positively charged beam of momentum $P=75$ MeV/c  was collected at 135$^{\circ}$ and transported in vacuum in the
 upgraded M13 channel\footnote{The M13 beamline at TRIUMF was modified in order to
 reduce the positron contamination in the beam.} \cite{M13paper} to the PIENU  apparatus. 

The PIENU apparatus, schematically shown in Figure \ref{detview}, was placed immediately
downstream of the exit of the M13 beamline. The end of the vacuum pipe was 
covered 
with a 101.6 ~mm diameter, 72.6 $\mu$m-thick stainless steel window.
To reduce the neutral background coming from the pion production target
 area and collimators in the beamline, a 24.8 cm-thick 2.57 $\times$ 2.29 m$^2$ (width $\times$ height)
 steel wall with a hole (26.67 cm diameter) for the beam
 pipe to go through was installed upstream of the detector assembly.

\begin{table*}[htbp]\footnotesize
\begin{center}
\caption[The PIENU Detector parameters]{PIENU detector parameters.}
\label{det_param}
\vspace{0.5cm}
\begin{tabular}{ lccccc }
\hline
\hline
\multicolumn{6}{ c }{\bf{Plastic scintillator counters}} \\
\hline
Trigger Counters & B1 & B2 & B3 & T1 & T2 \\ \hline
Size in X (Inner radius)  & 100~mm  & 45~mm  & 70~mm  & 80~mm  & (0)~mm  \\
Size in Y (Outer radius)  & 100~mm  & 45~mm  & 70~mm  & 80~mm & (171.45)~mm \\
Size in Z  & 6.604~mm  & 3.07~mm  & 8.05~mm  & 3.04~mm  & 6.6~mm  \\
Z position   &  -39.03~mm  & -30.02~mm  & 0~mm  & 19.92~mm  & 72.18~mm  \\
Photomultiplier model/ & H3178-51 & 83112-511 & XP2262B & 83112-511 & H3165-10\\
manufacturer & Hamamatsu & Burle & Photonis & Burle& Hamamatsu\\
Photocathode diameter  & 34~mm  & 22~mm & 44~mm  & 22~mm & 10~mm \\
\hline
\hline
\multicolumn{3}{l}{Veto Counters} & V1 & V2 & V3\\
\hline
\multicolumn{3}{l}{Inner radius } & 40~mm  & 107.95~mm  & 177.8~mm \\
\multicolumn{3}{l}{Outer radius } & 52~mm  & 150.65~mm  & 241.3~mm \\
\multicolumn{3}{l}{Size in Z } &3.175~mm  & 6.35~mm  & 6.35~mm \\ 
\multicolumn{3}{l}{Photomultiplier model/} &H3164-10  &  \multicolumn{2}{c}{H3165-10 }\\
\multicolumn{3}{l}{Photomultiplier manufacturer} &Hamamatsu  &  \multicolumn{2}{c}{Hamamatsu }\\
\multicolumn{3}{l}{Photomultiplier Photocathode diameter } &8~mm  & \multicolumn{2}{c}{10~mm }\\
\hline
\hline
\multicolumn{6}{ c} {\bf{Tracking detectors}}\\
\hline
\multicolumn{3}{l}{ Multi Wire Proportional Chambers} & WC1 & WC2 & WC3 \\
\hline
\multicolumn{3}{l}{Wire spacing } & \multicolumn{2}{c}{0.8~mm } & 2.4~mm  \\
\multicolumn{3}{l}{Number of Planes/wires/readout channels} &\multicolumn{2}{c}{ 3/120/40} & 3/96/48 \\
\multicolumn{3}{l}{Active area diameter } & \multicolumn{2}{c}{96.0~mm } & 230.4~mm  \\ 
\multicolumn{3}{l}{Cathode to Anode spacing } &\multicolumn{2}{c} {1.6~mm } & 2.0~mm  \\ 
\multicolumn{3}{l}{Wire diameter } & \multicolumn{3}{c}{15~$\mu$m }\\ 
\multicolumn{3}{l}{Wire orientation}& \multicolumn{3}{c}{0$^\circ$, +120$^\circ$, -120$^\circ$ }\\
\hline
\multicolumn{5}{l}{Silicon Strip Detector Pair (X and Y oriented strips)}& S1/S2/S3\\
\hline
\multicolumn{5}{l}{Active area }& 61 $\times$ 61~mm$^2$ \\
\multicolumn{5}{l}{Silicon strip pitch}& 80~$\mu$m \\
\multicolumn{5}{l}{Effective pitch after binding 4 strips }& 320~$\mu$m \\
\multicolumn{5}{l}{Number of planes/readout channels per plane} & 2/48\\
\multicolumn{5}{l}{Thickness (size in Z) }& 0.285~mm \\
\multicolumn{5}{l}{Separation between X and Y strip detectors }&  12~mm \\
\hline
\hline
\multicolumn{6}{c}{\bf{Electromagnetic calorimeter}}\\
\hline
\multicolumn{4}{l}{Crystal} & NaI(T$\ell$) & CsI \\
\hline
\multicolumn{4}{l}{Number used} & 1  & 97 \\
\multicolumn{4}{l}{Energy Resolution (FWHM) at 70 MeV} & 2.2\%  & $\sim$10\% \\
\multicolumn{4}{l}{Thickness (size in Z) }  & 480~mm  &  250~mm  \\
\multicolumn{4}{l}{Outer Radius } & 480~mm  &  - \\
\multicolumn{4}{l}{Approximate width $\times$ height for pentagon shaped CsI crystals} & -  & 90$\times$ 80~mm$^2$ \\
\multicolumn{4}{l}{Number of PMTs per crystal} & 19 & 1 \\
\multicolumn{4}{l}{Hamamatsu PMT model (central PMT for NaI(T$\ell$) was R1911-07)} & R1911 & R5543 \\
\multicolumn{4}{l}{Photomultiplier Photocathde diameter } & \multicolumn{2}{c} {76.2~mm } \\

\hline
\end{tabular}

\end{center}
\end{table*}

During normal data taking, the beam
was composed of 84\% $\pi^+$, 14\% $\mu^+$ and 2\%  $e^+$. It was imaged
by a circular multiwire proportional chamber package (WC1 and WC2, containing three wire planes each).
 Following   WC2, the beam was degraded by two plastic scintillator counters
 B1 and B2 (beam counters)
used for measuring time and energy loss for particle identification.
The beam counters were followed by two pairs of Si strip detectors (S1 and S2) with strips oriented along the
X and Y axes\footnote{A right-handed coordinate system is used with the origin at the center of 
the target counter. The Z-axis points in the beam direction, and the Y-axis 
points upwards.}. An 8 mm thick plastic scintillator target (B3) 
was sandwiched between S2 and another X-Y Si strip pair (S3). A circular multiwire proportional chamber
 WC3, containing
 three wire planes, was positioned downstream of S3 and
 sandwiched between two plastic scintillators, T1 and T2 (telescope counters).

Incoming pions stopped in the center ($\pm 1$~mm along the Z-axis) of B3, and then
decayed at rest. 
 A 
 coincidence of signals from T1 and T2 was used to define the 
 on-line positron acceptance.   WC3 and S3 were
 used to reconstruct charged particle  tracks 
and define the final acceptance off-line.
The 15\% solid-angle acceptance of the crystal spectrometer  was defined by a radius cut (r=6 cm) 
 in the central plane of WC3, which was located 5.5~cm downstream\footnote{Distance from the center of B3 to the central plane of WC3} of B3.
Accepted decay products entered a large 
 single crystal NaI(T$\ell$) detector (48~cm long and 48~cm diameter) \cite{LEGS:1997}. 
In order to absorb shower leakage from 
the NaI(T$\ell$) crystal and to detect large angle photons from radiative \pienu$\gamma$ decay,
 it  was enveloped by four rings of pure CsI crystals \cite{Adler:2008zza}.
 The NaI(T$\ell$) and CsI crystal array served as the calorimeter to
 measure the energy of the decay product positrons.

\begin{figure}[htbp]
\begin{center}
\includegraphics[trim=0cm 0cm 0cm 0cm, angle=0, clip=true,width=0.8\columnwidth]{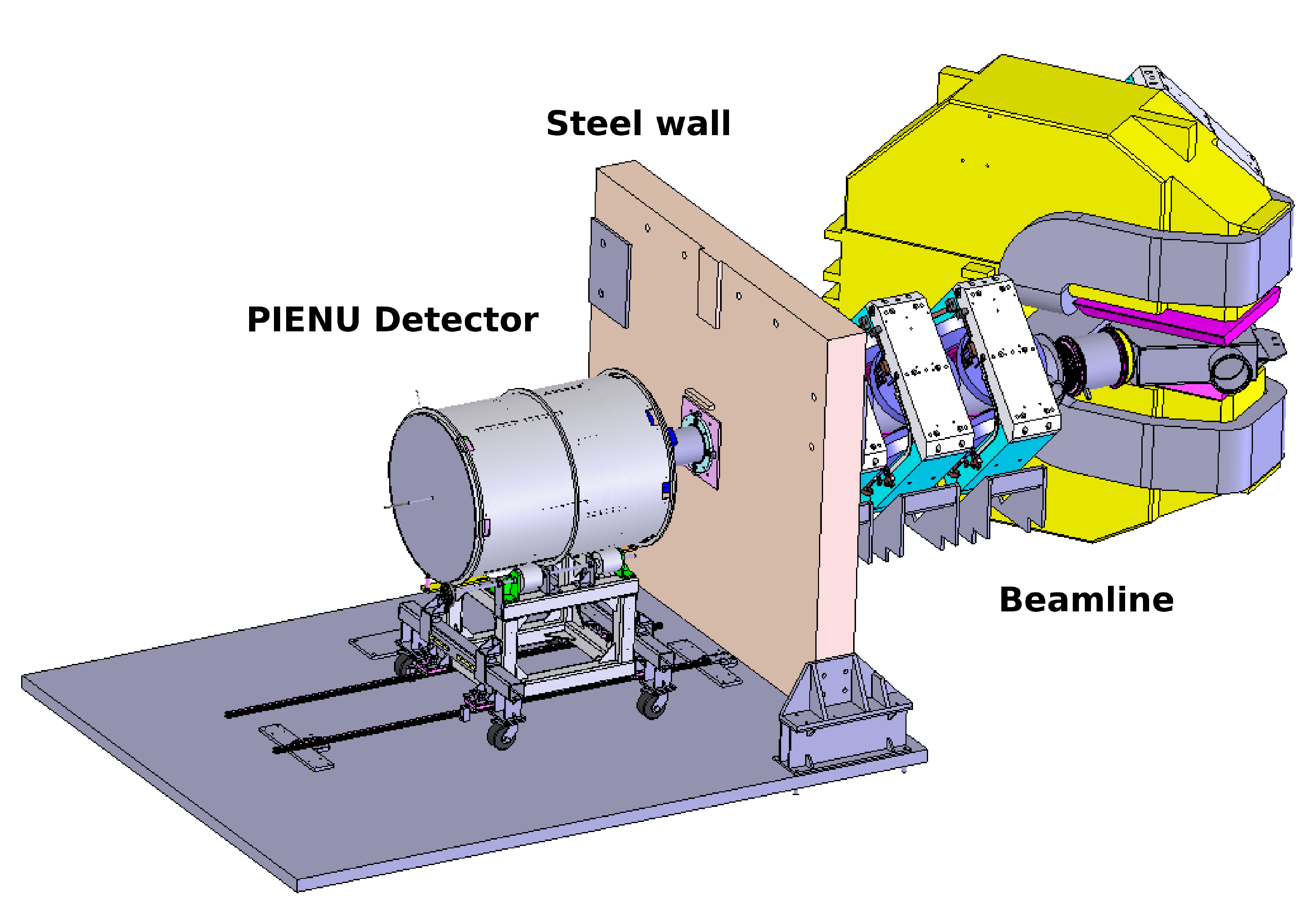}
\caption{ Computer generated rendering of the PIENU detector and the steel wall used as
 passive shielding at the end of the M13 beamline.}
\label{fig:pienu_assy}
\end{center}
\end{figure}

Three annular scintillators, which covered
 inactive material, were used for systematic studies: V1 covered the front frame
 of WC1, and V2 and V3 covered the
 frame of WC3 and the front flange of the NaI(T$\ell$) crystal enclosure, respectively.
Detector parameters are given in Table \ref{det_param}.

\section{Design and construction}

The experimental apparatus was constructed in modules (PIENU-I and PIENU-II) for simplification of installation and servicing.
PIENU-I consisted of V1, WC1, WC2, B1, B2, S1, S2, B3, S3 and T1, while PIENU-II included WC3, T2, V2, V3, NaI(T$\ell$) and CsI.

PIENU-I was attached to the end of the beam pipe 
and could be divided into three sub-assemblies (V1+WC1+WC2, B1+B2+S1+S2 and B3+S3+T1).
The PIENU-II detector assembly was supported on a cart.
During normal running conditions the cart rested on rails which were aligned 
with the beamline and PIENU-I in the X-Y plane.
Modularity allowed a number of supplemental measurements to be performed aimed at evaluating systematic effects
during data taking. One of the measurements was determination of the response of the
NaI(T$\ell$) and CsI crystal array to a beam of 70 MeV/c incident positrons at various positions and
 angles of entrance as described in Section \ref{sec:cal_performance}.
 This measurement was made possible by leaving only the first sub-assembly of PIENU-I in place, 
containing V1+WC1+WC2, which detected the incoming beam, and removing the rails in order to allow the  
PIENU-II assembly to be moved freely within the experimental area.

 In 2010, the experimental area was enclosed in a temperature-controlled tent in order
 to make temperature-related changes in detector's performance negligible. The temperature inside the tent
 was kept at 20$^\circ \pm 0.5^\circ$~C.

\section{Scintillation counters}
\label{sec:scint}

The plastic scintillators were fabricated from Bicron BC-408 
(polyvinyltoluene) scintillator \cite{BC408} chosen for
 its high light-output, fast rise (0.9 ns), and short decay (2.1 ns) times.
 The B1 beam scintillator was large enough to fully cover the aperture
 of the multiwire chamber package WC1 and WC2. 
The B2 scintillator, which was smaller than the B3 scintillator,
 ensured that the particles going through B2 would hit B3\footnote{The rectangular-shaped B3 and 
T1 counters were rotated by 45$^o$ around the beam axis
 due to spatial limitations of the mechanical assembly
 which supported the scintillator along with their respective 
light guides,  photo multiplier tubes (PMTs),
 and a number of other subdetectors (WC1, 
WC2, S1, S2, S3).}.
 T1, located downstream of B3, defined the positron timing. 
Due to space restrictions and their circular shape,
T2 and the veto scintillators (V1-V3) were read out with 1~mm diameter  
wave-length-shifting (WLS) fibers (Kuraray Y-11) 
embedded in  grooves (1.1$\times$1.1~mm$^2$ in cross-section) machined on the scintillator surfaces;
 groove spacing was 10~mm, 3~mm, 10~mm and 11~mm for T2, V1, V2 and V3\footnote{T2 had linear parallel grooves traversing the counter's surface, while
annular veto counters V1,V2 and V3 had circular grooves.} counters respectively.

\begin{figure}[!ht]
\begin{minipage}[!ht]{0.48\columnwidth}
\centering
\includegraphics[trim=0cm 0cm 0cm 0cm, clip=true,angle=0,width=0.95\linewidth]{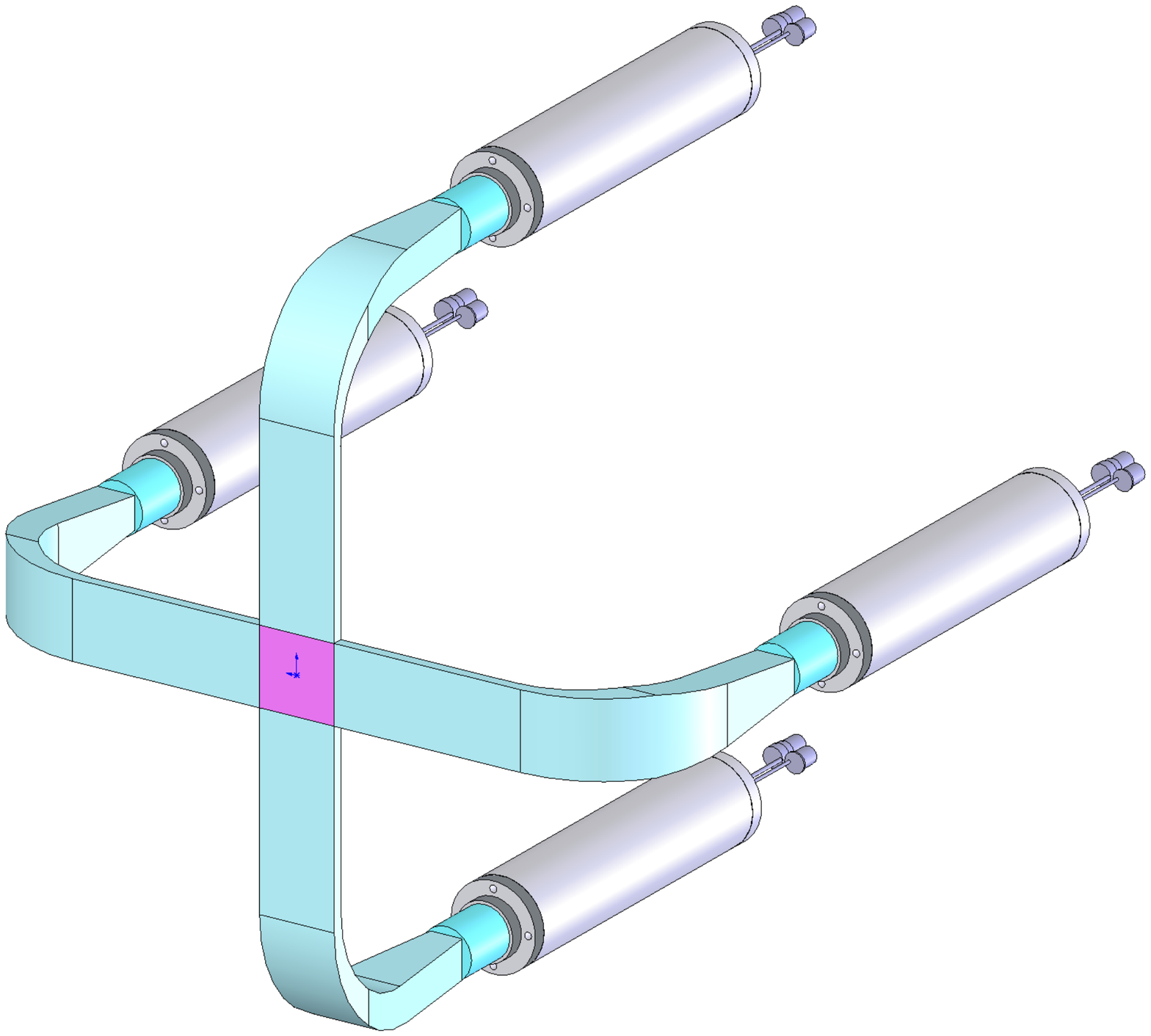}
\caption{Schematic view of the B1, B2, B3 and T1 scintillator readout. Light guides were bent adiabatically and coupled to PMTs through acrylic cylinders.}
\label{scint:readout_scheme}
\end{minipage}
\hspace{0.04\columnwidth}
\begin{minipage}[!ht]{0.48\columnwidth}
\centering
\includegraphics[trim=0cm 0cm 0cm 0cm, clip=true,angle=0,width=0.95\linewidth]{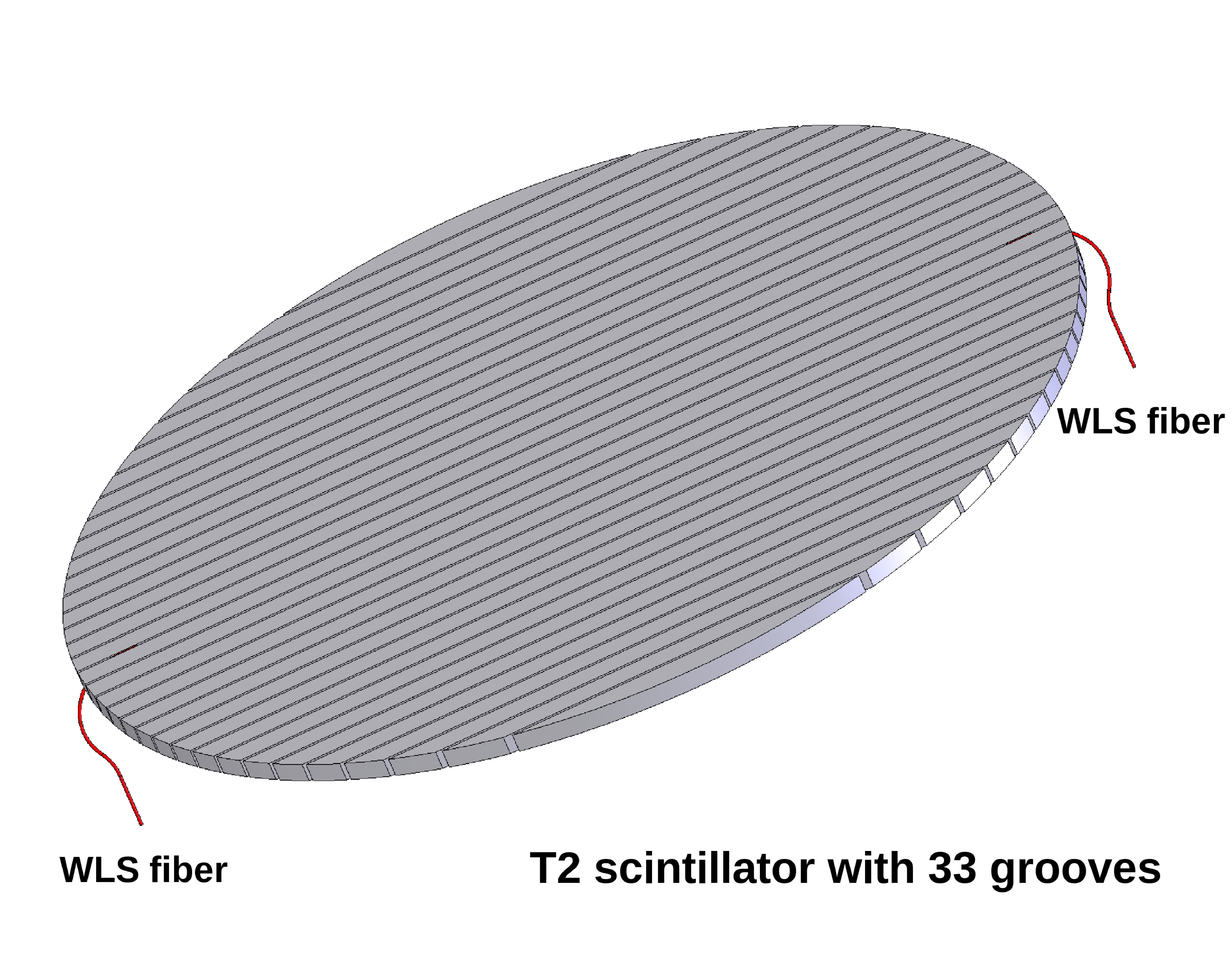}

\caption{Schematic drawing of the T2 scintillator. 33 parallel grooves ($1.1\times1.1 ~mm^2$) machined on the surface of the counter were made deeper towards the edge to allow the WLS fibers to exit perpendicular to the counter's surface (see text for details). One such readout WLS fiber is shown for illustration.}

\label{scint:t2readout}
\end{minipage}
\end{figure}

 Collection of the scintillation light for all the rectangular-shaped scintillators
was done with adiabatic
light guides which were cut from UV transparent acrylic sheets and polished.
It was important to have a uniform 
light collection efficiency 
in order to
maximize energy resolution. A schematic of the four-way
 readout configuration is shown in
 Figure \ref{scint:readout_scheme}. Simulations of the optical transport with Detect2000 \cite{Detect2000} were used to predict
the uniformity of the light collection in this configuration, and were later 
confirmed with measurements. For example, a variation of $<$ 1.2$\%$ was observed in the visible energy deposited by beam muons in B1 over the counter's surface.

\begin{figure}[!ht]
\begin{minipage}[!ht]{0.48\columnwidth}
\centering
\includegraphics[trim=0cm 0cm 0cm 0cm, clip=true,angle=0,width=0.9\linewidth]{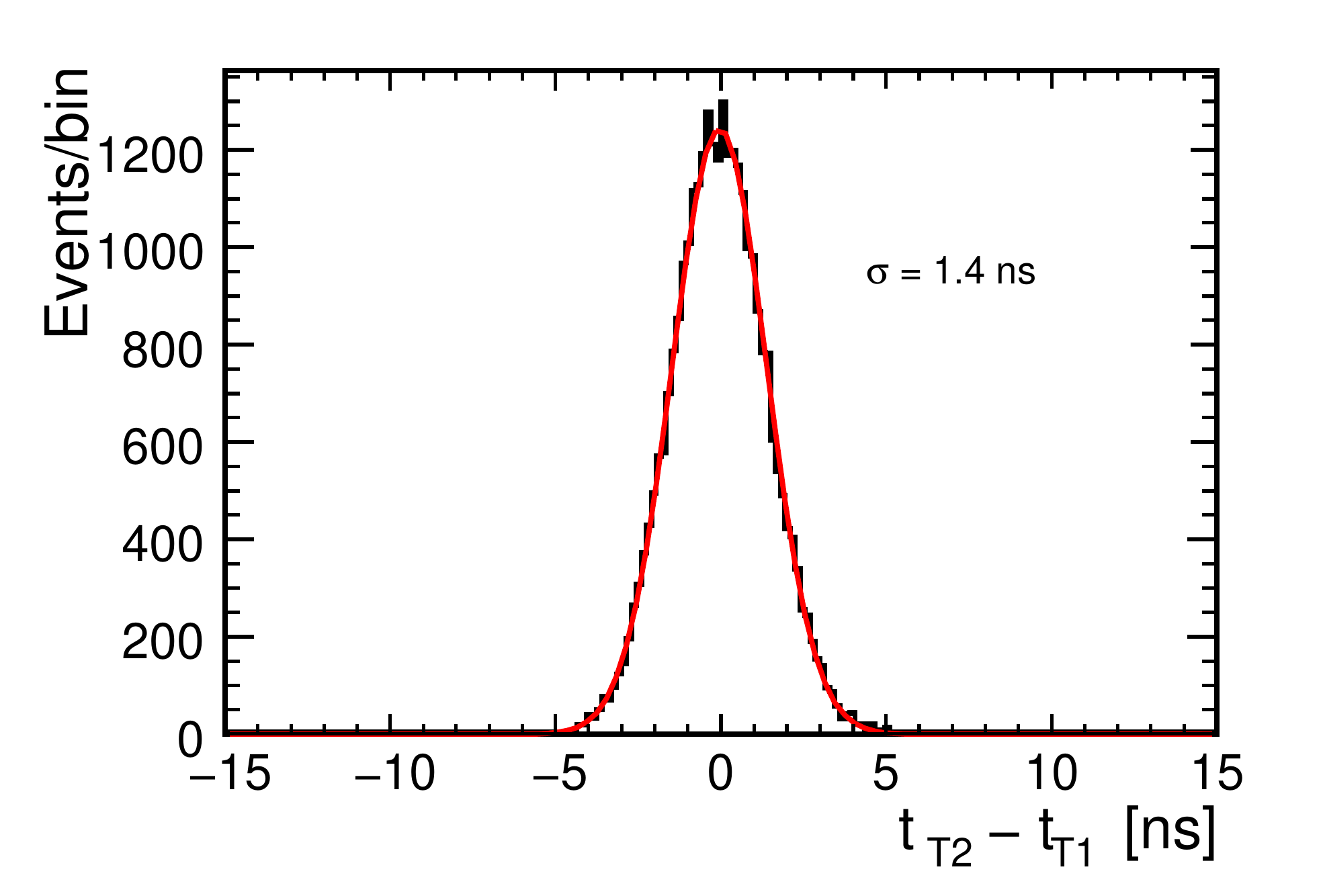}
\caption{The observed time difference between T2 and T1. The T1  time is determined by fitting a waveform, while
the T2 time is defined as the time of the earliest pulse found within the 
waveform. Due to  WLS fiber readout of T2 several pulses could be seen for one particle. Histogram is data, and the smooth line is a Gaussian fit.}
\label{fig:t2_res}
\end{minipage}
\hspace{0.04\columnwidth}
\begin{minipage}[!ht]{0.48\columnwidth}
\centering
\includegraphics[trim=0cm 0cm 0cm 0cm, clip=true,angle=90,width=0.9\linewidth]{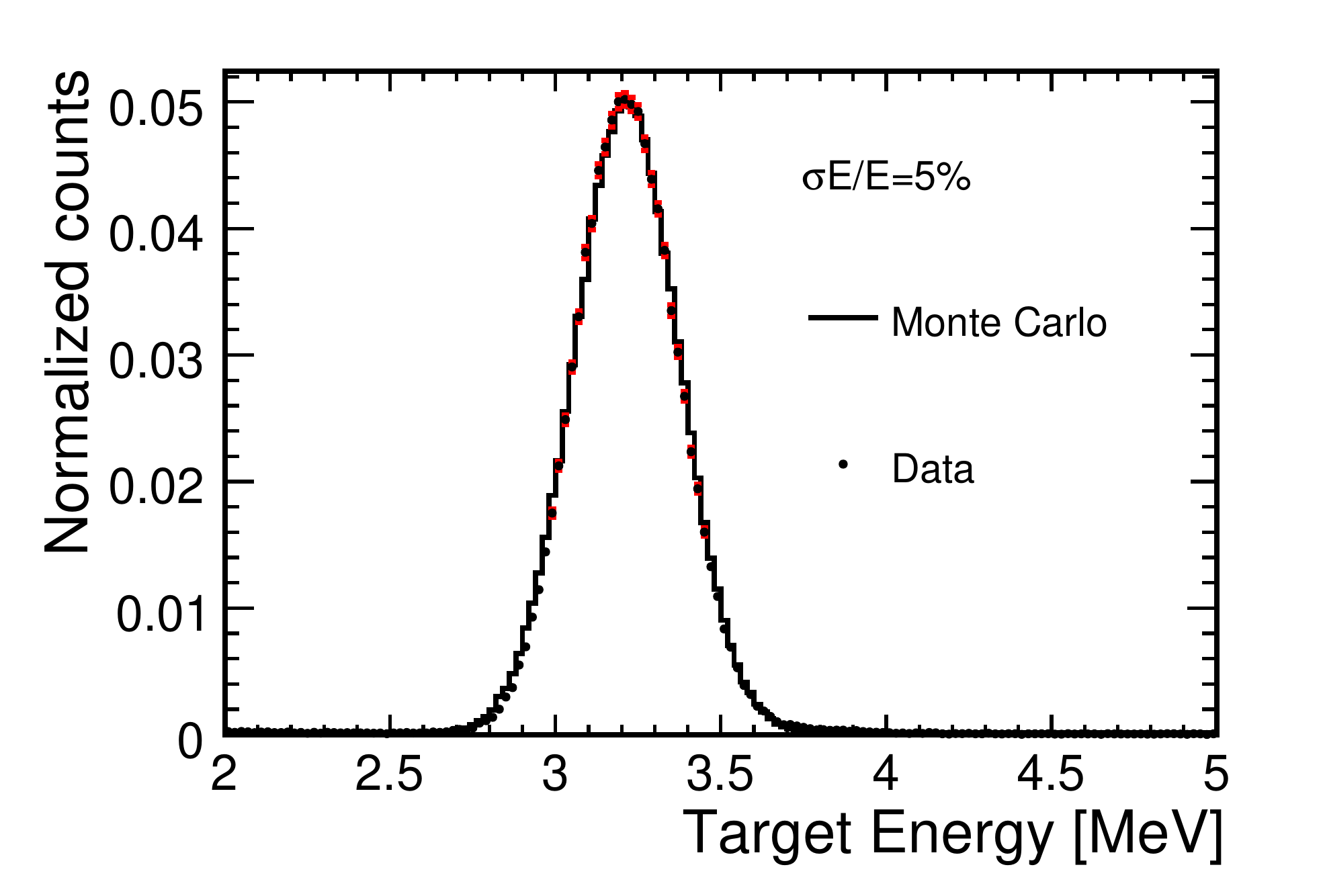}
\caption{Visible energy deposited in  B3  by a  $\mu^+$ with kinetic energy of 4.1 MeV from $\pi^+ \rightarrow \mu^+ \nu_{\mu}$  decay. 
The data is represented by points, and a  Monte Carlo simulation by the 
histogram. The mean visible energy is approximately 3.2 MeV due to saturation of 
scintillation light.} \label{fig:tg_res}
\end{minipage}
\end{figure}
Each scintillator was read out with four photo-multiplier tubes (PMTs), except for V1, V2 and V3 which
 were each read out with a single PMT. 
PMTs were selected by photocathode size and ability to match either the 425~nm wavelength of
 the maximum emission of the Bicron-408 or
 the 476~nm wavelength of the emission peak of the WLS fibers. Information on the scintillator size and PMTs used is listed in Table \ref{det_param}.
All plastic scintillator PMTs were read out with 500 MHz ADCs (see Section \ref{sec:daq}).

T2 was the largest scintillator read out with WLS fibers, having 33 parallel grooves
 machined on the downstream side.
 Near the edge of the counter, the grooves were made gradually deeper
 so that the WLS fibers lying in the grooves could exit perpendicular
 to the counter's surface with minimum transition space and point upstream where they could be bundled
 and read out by PMTs (see Figure \ref{scint:t2readout} for schematic
of the T2 grooves). The minimum radius of the groove 
 curvature at the edges of T2 was 19.5~mm. Fibers were thermally pre-bent to reduce 
stress.
T2 fibers were read out on both ends, and bundled into two groups (1-16 and 17-33).
 As a result, T2 was read out using four PMTs.
It had a time resolution of $\sigma_t=$1.4~ns for minimum ionizing particles when measured
 against T1 (see Figure \ref{fig:t2_res}) and its efficiency for positrons was measured
 to be greater 
than 99.7\% within the acceptance region.

The most critical scintillation counter, B3, had an energy resolution
 of 5\% r.m.s. ($T_{visible}= 3.22 \pm 0.16$ MeV) for  4.1  MeV $\mu^+$  
from $\pi^+\rightarrow\mu^+\nu_{\mu}$ decay
(see Figure \ref{fig:tg_res}). The time difference between T1 and B1 was used to determine the decay
 time with a resolution of $\sigma_t = $138~ps for beam positrons. 
 Both T1 and B1 times were determined from fits of 500 MHz-sampling-frequency waveforms.\\

\section{Calorimeter}

The purpose of the PIENU calorimeter system was to detect 
decay positrons
and photons from the \pienu and \pienu$\gamma$ decays as well as
 from the 
 $\mu^+\rightarrow e^+\nu_e \bar{\nu}_{\mu}$ and $\mu^+\rightarrow e^+\nu_e \bar{\nu}_{\mu}\gamma$~ following the decay $\pi^+\rightarrow\mu^+\nu_{\mu}$. 
It was composed of a large monolithic NaI(T$\ell$) crystal surrounded by 97 pure
CsI crystals. Scintillation light collection was realized through directly coupled PMTs.

\subsection{NaI(T$\ell$) crystal}
The main calorimeter element was a single crystal of thallium-doped NaI 
on-loan from Brookhaven National Laboratory 
\cite{LEGS:1997}.
The crystal was enclosed in a 3~mm thick 
aluminum enclosure which had nineteen 76.2~mm diameter circular quartz windows at
 the rear end.
The aluminum front face of the NaI(T$\ell$) was 0.5~mm thick.
Each window was viewed by a 76.2~mm diameter Hamamatsu R1911 PMT (with the exception of the central PMT 
R1911-07) surrounded by a $\mu$-metal shield.

The surface of the crystal was covered with reflective material. An optical simulation was performed using Detect2000 \cite{Detect2000} to study the dependence of the energy deposited in the crystal on the entrance location of
the particle.  
Light emitted by the crystal was uniformly reflected\footnote{For the simulation both specular and diffuse reflectors were modeled, with the reflection coefficient varying between 0.95 to 0.99 and a refractive index of 1.85 was assumed for the NaI(T$\ell$) crystal.}, and 
thus a similar amount of light was seen by individual PMTs independent
 of the entrance position of the ionizing particle on the 
front face of the NaI(T$\ell$) crystal. This was confirmed within 2\% by bench tests using 511~KeV and 1274~KeV $\gamma$ rays from a $^{22}$Na radioactive source.
The calibration of the NaI(T$\ell$) crystal was done with 75 MeV/c beam positrons using  a dedicated trigger.

\subsection{CsI crystal array}
Each crystal within the CsI array was 25~cm (13.5 radiation lengths) long and had 
pentagonal shape, with average width of 9~cm and height of 8~cm (4.5 radiation lengths). The 
crystals were arranged to form two upstream and two downstream
 concentric layers around the
NaI(T$\ell$). Each concentric layer  of crystals was supported by a 2~mm thick stainless steel cylinders with a 2-mm thick fin separating and supporting every 3-5 crystals from its neighbors (6 fins per layer).
 The resulting array of CsI crystals was 50~cm long and 16~cm (9 radiation lengths) thick in the radial direction. 
The CsI array was continuously
 flushed with nitrogen gas to keep the humidity level low.
Each crystal was read out by a fine-mesh 76.2~mm diameter Hamamatsu R5543 PMT \cite{Komatsubara}\footnote{The components 
of the magnetic background in the experimental area have been measured to be $<$2 Gauss at the location of the detector, 
well within the operational specifications of these fine-meshed PMTs. Before being brought to TRIUMF, the 
crystals and their PMTs were used as endcap photon-veto detector in the E949 experiment at BNL \cite{Chiang:1995ar}.}
The pure CsI pulse shape has two components: fast and slow with 30~ns and 680~ns decay times, respectively.
The fast component constitutes about 20\% of the total pulse.
In order to suppress the slow component, a UV-transmitting optical filter was present in front of each PMT.
Each crystal had a YalO$_3$:Ce$^{245}$Am source \cite{YalO} attached to its front face to monitor the crystal's light output
 and the PMT's gain. The source emits light at a frequency of 50~Hz with similar wavelength and pulse width as the
 CsI scintillation light with an equivalent
energy deposit of $10$~MeV.
Each crystal was also connected via a quartz fiber to the output of a Xe lamp whose flash was triggered at 2 Hz during
 data taking. This Xe lamp system was implemented to monitor CsI PMT's gains.
Seven reference PMTs of the same type as the ones used for the calorimeter were enclosed in an incubator maintained at a constant 
temperature of 24.0 $^{\circ}$C. The Xe lamp was enclosed in an identical incubator at the same temperature. The Xe-lamp was 
connected to the reference PMTs through the same system as the CsI crystals so that any changes in the Xe-lamp output could 
be tracked by the reference PMTs\footnote{The Xe lamp system was found to be stable to less than 1\% level and was used for monitoring. Since
the whole experimental apparatus was enclosed in the temperature controlled tent, no temperature-dependent variations in the PMT gains 
were observed, while the PMT gain degradation over time was on the order of $\approx$1\% per year.}.

The energy calibration of the CsI array was performed with cosmic rays. A dedicated trigger, based on coincidence of the two concentric
CsI rings, provided a sample of cosmic ray data. The energy peak due to minimum ionizing muons going through the crystals was compared
with a detailed Monte Carlo (MC) simulation of the system to obtain the energy calibration.
\begin{figure}[htbp!]
\begin{center}
\includegraphics[trim=0cm 0cm 0cm 0cm, clip=true,width=\linewidth]{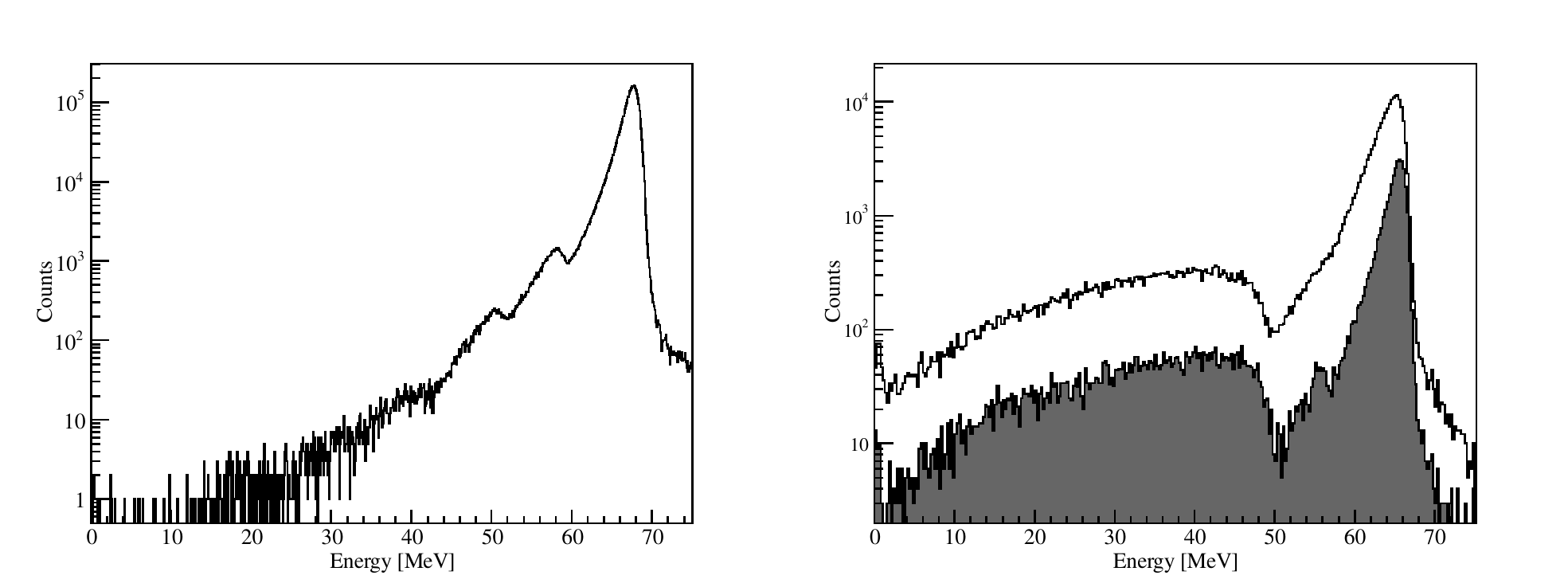} 
\caption{{\bf Left:} Response of the NaI(T$\ell$) crystal to a positron beam of momentum 70 MeV/c. Below the full energy peak, the additional low energy peak structures are due to
photoabsorption followed by neutron escape from the crystal \cite{phn}. {\bf Right:} \pienu$ $decay spectrum with additional cuts
for suppressing the \pimue $ $ decays, which can be seen in the energy spectrum up to approximately 50 MeV. The first 
structure due to photoabsorption in the crystal becomes visible to the 
left of the main peak (from the \pienu $ $decay positron at around 65 MeV) as one restricts the acceptance.
The shaded histogram represents a tighter-than-nominal 40~mm radial acceptance cut 
and the unfilled histogram represents the nominal acceptance cut of 60~mm.}
\label{photonuclear}
\end{center}
\end{figure}
\subsection{Performance}
\label{sec:cal_performance}
{\bf NaI(T$\ell$):} the response of the NaI(T$\ell$) crystal was investigated
by reconfiguring the detector to include PIENU-II and only the first sub-assembly of PIENU-I (V1, WC1, and WC2).

The response of the
crystal to a 70~MeV/c positron beam entering its front face at the center is shown in Figure~\ref{photonuclear} (left).
The high energy peak corresponds to the full energy of the beam positrons. After deconvoluting the beam momentum width
of $0.5$\%, the resolution at 70~MeV/c was 2.2\% (FWHM). Below the beam energy, additional structures are visible \cite{phn} due
to nuclear photoabsorption and subsequent neutron escape.
In this process, one or more photons in the shower initiated by the positron are
absorbed by iodine and neutrons emitted after the absorption escape the crystal reducing
the measured energy by the neutron binding energy of approximately 8 MeV. 
Structures corresponding to up to three escaping neutrons can be seen in Figure~\ref{photonuclear} (left).
 This mechanism was consistent with a MC simulation of the crystal response including electromagnetic and
hadronic interactions  \cite{phn}.

\begin{figure}[htbp!]
\begin{center}
\includegraphics[trim=0cm 0cm 0cm 0cm, clip=true,width=0.8\linewidth]{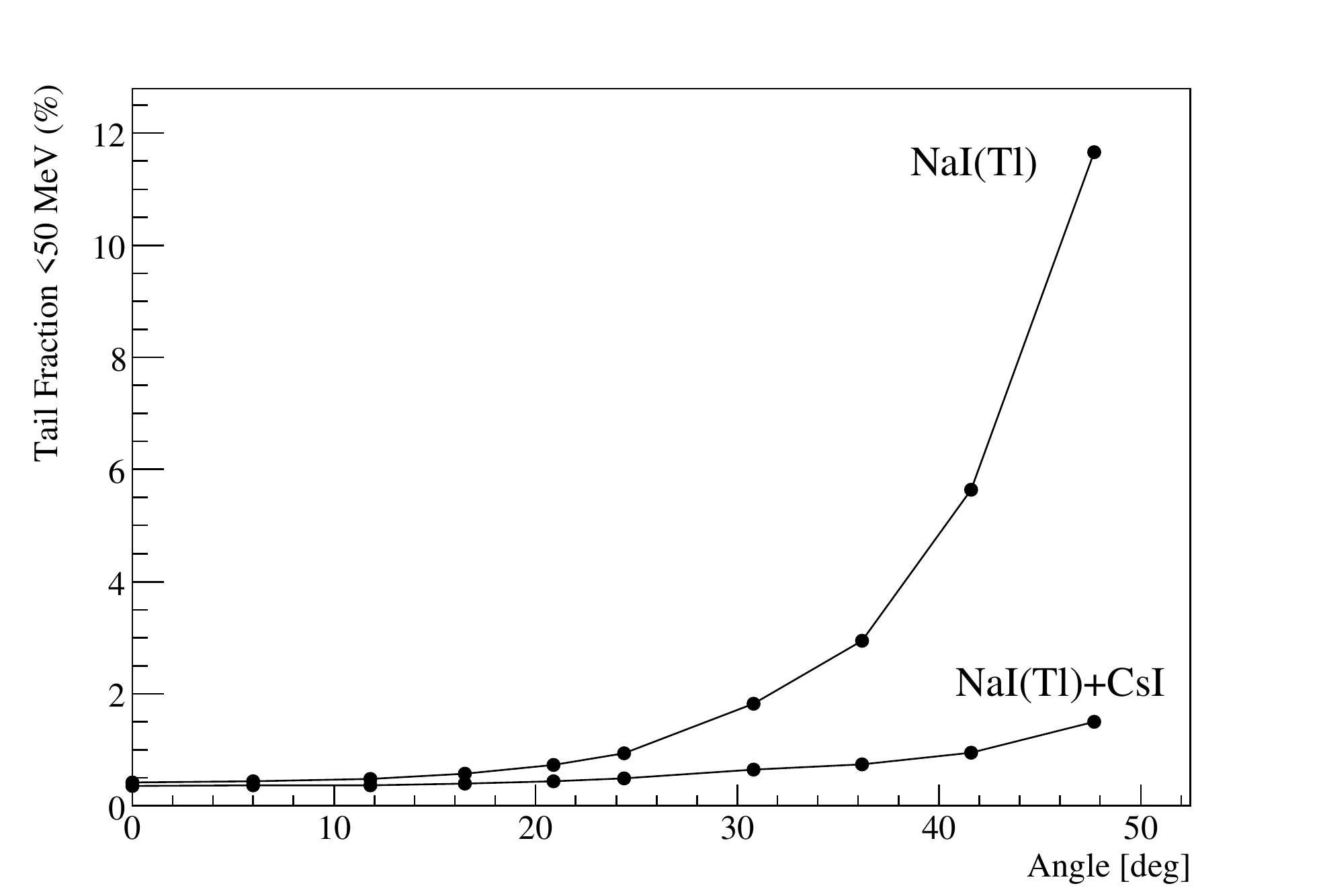}
\caption{Fraction of events below 50~MeV for the NaI(T$\ell$) crystal alone and the NaI(T$\ell$) plus CsI combination as a function of the angle between the beam and the calorimeter horizontal axis . Error bars
 are too small to be visible.}
\label{tailfraction}
\end{center}
\end{figure}

Photonuclear  effects are also present when measuring the \pienu decay positrons. 
 An energy spectrum of the \pienu decay positrons with a small ($<$15\%) contamination of positrons from the 
 $\pi \to \mu \to e$  decay chains  is shown in Figure~\ref{photonuclear} (right), where 
the presence of a structure due to one-neutron escape below the main peak (see also \cite{mnu}) is evident.

{\bf CsI Array:} The purpose of the CsI array was the collection of 
the electromagnetic shower leakage from the main NaI(T$\ell$) calorimeter.
 A critical 
part of the PIENU experiment was the reduction
and measurement of the low energy tail from the \pienu
 decay to reduce systematic uncertainties in the correction for events buried under the $\pi\to\mu\to e$ decay chain energy distribution. During the measurement of the calorimeter response 
a positron beam was used to investigate the fraction of the low energy tail as a function of the angle between the beam and the NaI(T$\ell$)
 crystal face. 
The crystal face was rotated to 10 different angles (up to 48$^o$) with respect to the beam. The results are shown 
in Figure~\ref{tailfraction} where the
fraction of events below 50~MeV (tail fraction) is reported as a function of the angle. When only the energy of the NaI(T$\ell$) crystal is
considered, the low energy tail fraction reaches $12$\% for the highest angle, while if the CsI energy is added, the fraction
never exceeds 2\% over the whole angular range.

\section{Tracking devices}

\begin{figure}[htbp]
\begin{center}
\includegraphics[trim=0cm 0cm 0cm 0cm, clip=true,angle=0,width=0.95\linewidth]{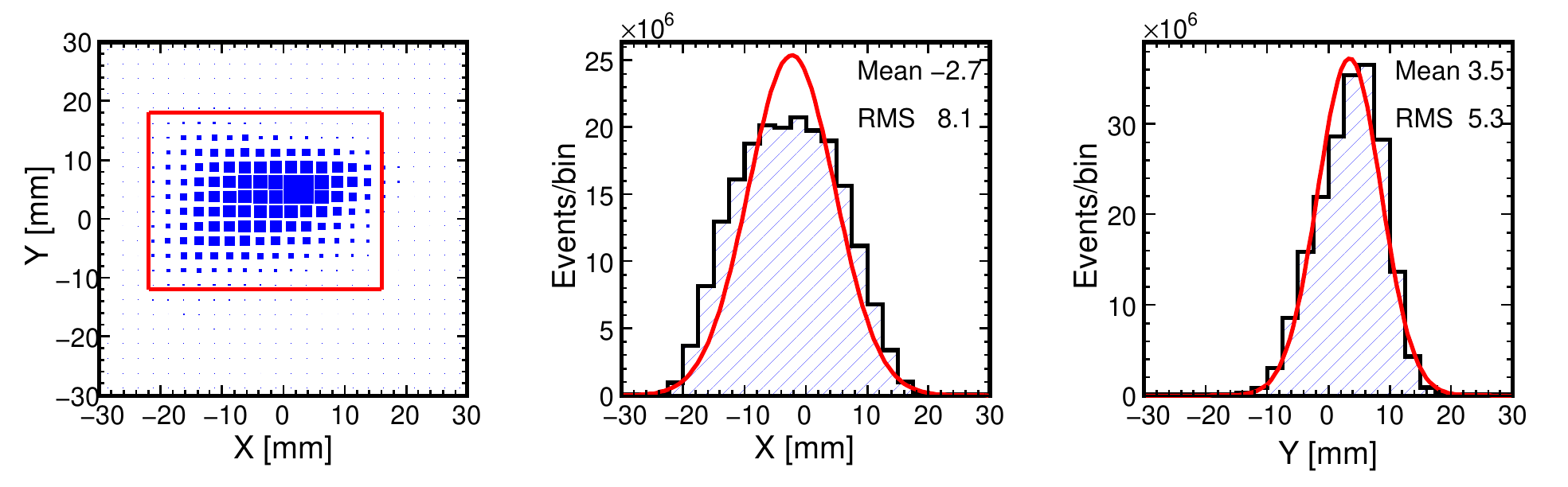}
\caption{Beam profile plots. Scatter plot (Y vs X) of the beam profile at the center of the WC1, the red box indicates the typical boundaries of the incoming particle selection cut (left plot). X  and Y profiles of the beam at the center of WC1 (center and right plots, respectively). X and Y coordinates are determined by reconstructing the 
incoming particle's track using WC1 and WC2.}
\label{fig:WC1}
\end{center}
\end{figure}

The beam wire chambers WC1 and WC2 were used to track the incoming pions  (see Figure \ref{fig:WC1}).
 Together with the two X-Y planes of silicon strip detectors (S1 and S2) they could detect pion decay-in-flight events upstream of the target (see \S\ref{S:perfo}). 
The downstream trackers (S3 and WC3) were used to track the decay positrons after the target. The reconstructed track's position at the center plane of WC3 was
used to define the acceptance of the positrons at the entrance of the calorimeters. 
Upstream (WC1, WC2, S1, and S2) and downstream (S3 and WC3) tracker information was also used to reconstruct the position of the pion's decay vertex in the target. 
 WC3 was also used to detect pile-up events which were not detected by the
 upstream trackers or S3 which had smaller recording time windows (see Section \ref{sec:daq}).

\subsection{Multiwire proportional chambers}
 Each wire chamber (WC1, WC2 and WC3) consisted of three wire planes which were rotated by an angle of
 120$^{\circ}$ with respect to each other. 
 The chambers were filled with a tetrafluoromethane (CF$_4$) - isobutane (C$_4$H$_{10}$) mixture (80\% - 20\%) at atmospheric
 pressure.
Each wire plane of WC1 and WC2 had 120 wires (0.8~mm in pitch) grouped by three. Each group was
 connected to a read-out pad. The number of read-out channels was 40 per plane. The active diameters of
 WC1 and WC2 were 9.6 cm.
The chamber WC3, mounted on the flange of the NaI(T$\ell$) crystal enclosure, measured the position of the
 decay positrons at the entrance of the calorimeter, and defined the acceptance region. Each plane had 96 wires with a pitch of 2.4~mm.
The wires were grouped in pairs which reduced the number of channels to 48. The active diameter of WC3 was 23~cm. The efficiency of every plane was 
measured to be 
$>$ 99\% for beam positrons.

\subsection{Silicon strip detectors}
\begin{figure}[htbp]
\begin{center}
\includegraphics[trim=0cm 0cm 0cm 0cm, clip=true,width=0.6\linewidth]{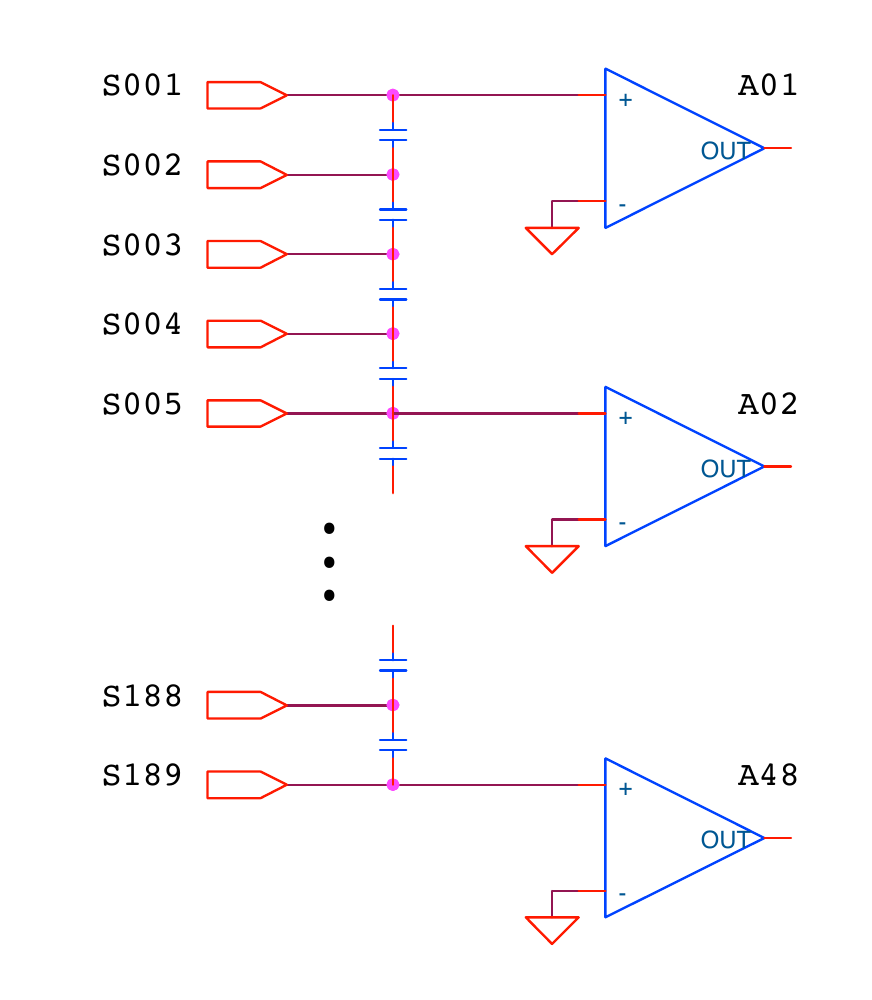}
\caption{Schematic drawing of the silicon strip detector readout scheme.}
\label{fig:Silicon}
\end{center}
\end{figure}

 Each set of silicon strip detectors (S1, S2, and S3) consisted of two identical hybrids fixed back to back with
 perpendicular orientation of strips (to measure X and Y coordinates). Each silicon detector had an active
 volume of $61\textrm{ mm} \times 61\textrm{ mm} \times 285~\mu\textrm{m}$. The Si sensor was a single sided
 AC-coupled micro-strip detector. In the original design \cite{SCT}, strips had a pitch of 80~$\mu$m but since the required 
resolution for the PIENU experiment was of the order of 300~$\mu$m, four strips were bound to one read-out line. For further reduction of the number of readout channels, the readout lines were interconnected with 
capacitors and only every fourth line was read out by an amplifier as shown in Figure \ref{fig:Silicon}.
Forty eight channels per silicon plane (288 channels in total) were read out. The capacitive network formed a
 charge division line where the reconstruction of the ionization amplitude and position was made by weighting 
of two or three channels that typically had a signal due to a charged particle traversing the sub-detector.
 The final position resolution (rms) provided by each silicon plane was 95~$\mu$m if at least two readout
 channels had a signal and 370~$\mu$m when the signal was recorded in only one readout channel. In order to reduce 
the data size,  a hardware threshold was set to suppress channels with no hits. In S1 and S2, the thresholds 
were optimized for the detection of pions while the S3 thresholds were set lower to ensure that the detection 
efficiency for decay positrons in at least one plane (X or Y) was higher than 99\%.\\
\subsection{Performance}\label{S:perfo}
\begin{figure}[htbp]
\begin{center}
\includegraphics[trim=0cm 0cm 0cm 0cm, clip=true,width=0.5\columnwidth]{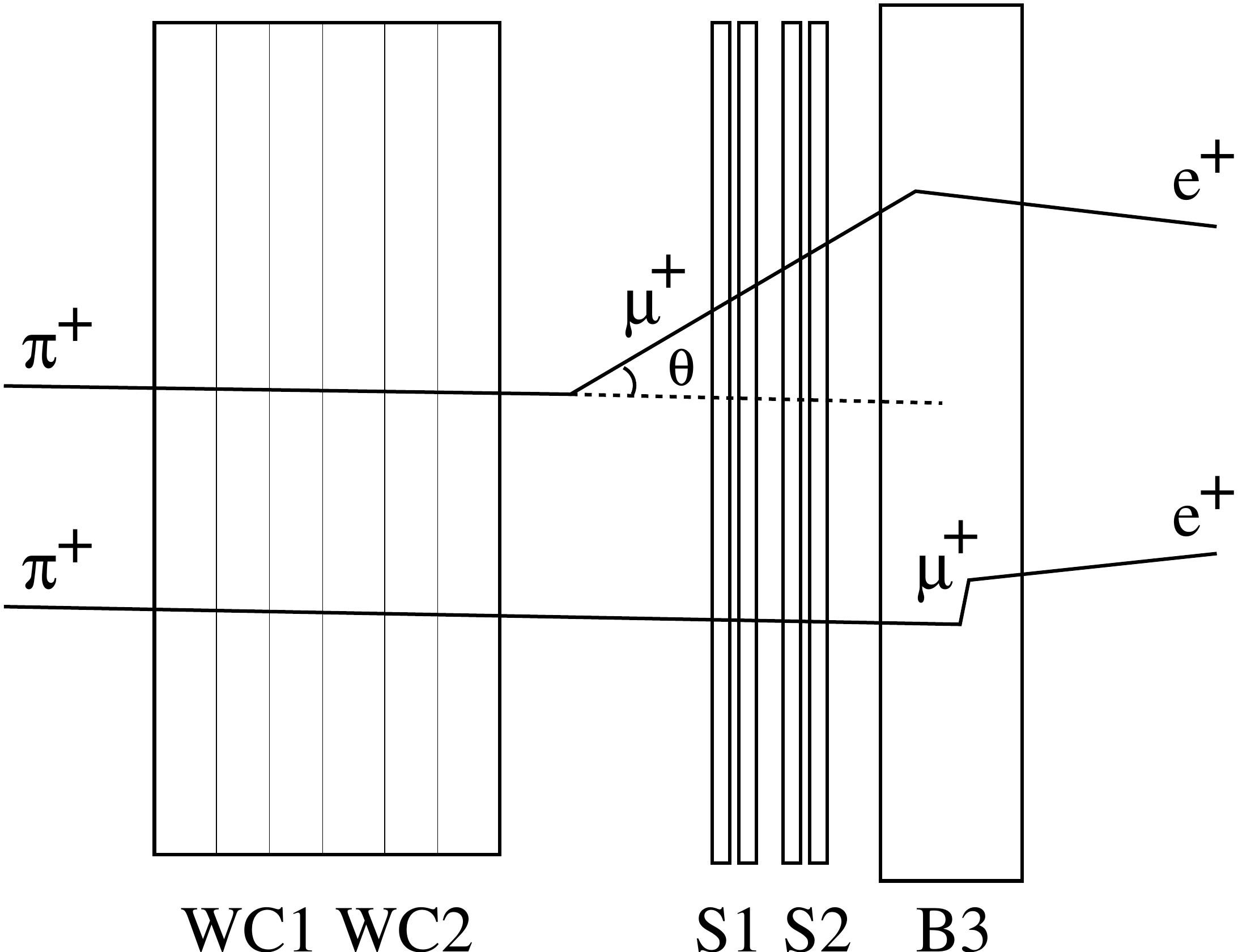}
\caption{Schematic representation of pion-DIF and pion-decay at rest (PDAR) event. Events with decay in flight show an inconsistency between the directions of the track segments reconstructed in WC1 and WC2, and S1 and S2, while track directions are consistent for decays at rest.}
\label{fig:PDIF}
\end{center}
\end{figure}

This compact assembly of tracking devices was designed to minimize the major source of background for the 
measurement of \pienu decay due to decay of pions in flight. 
About 3.6\% of the pions entering the PIENU detector\footnote{This result is obtained from the Geant4 simulation. A pion is denoted as entering the detector if it enters the first detector element WC1} decay in flight upstream or inside the target. 
This type of decay, labeled pion-DIF for pion decay-in-flight, will have a low energy deposit in the target 
and can therefore mimic the energy deposition of a \pienu decay.  About half of those pion-DIF events happen 
upstream of the target, between WC2 and S1, and can be identified by the tracking detectors as illustrated 
in Figure~\ref{fig:PDIF}. 

\begin{figure}[!ht]
\begin{minipage}[!ht]{0.48\columnwidth}
\centering
\includegraphics[trim=0cm 0cm 0cm 0cm, clip=true,angle=0,width=0.95\linewidth]{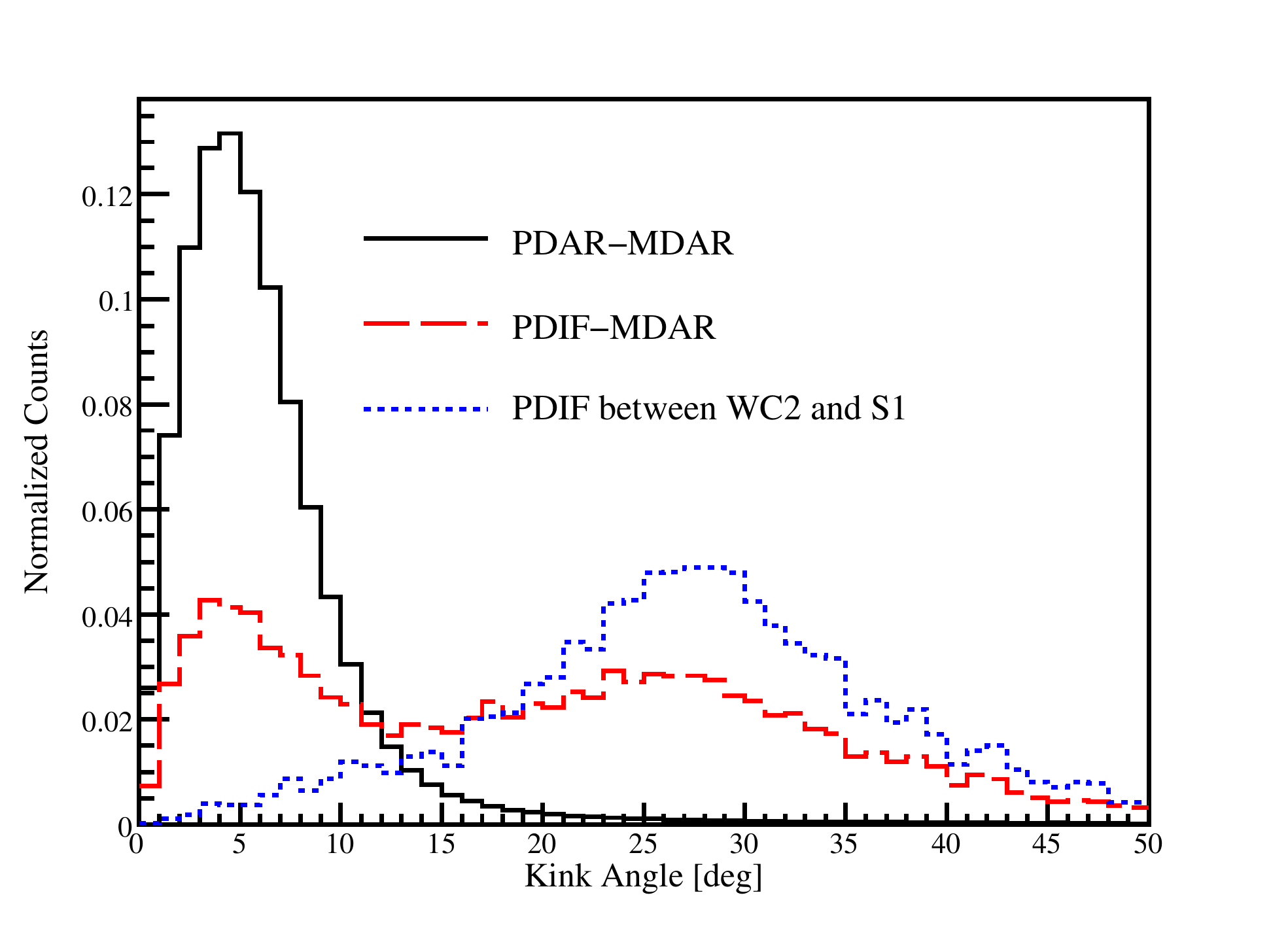}
\caption{Kink angle measured by WC1-WC2 and S1-S2 tracking devices from MC.
About half of the pion-DIF events happen after S1 and cannot be distinguished from pion-DAR (PDAR) or muon-DAR (MDAR) events. DAR stands for decay-at-rest.}
\label{fig:kinkMC}
\end{minipage}
\hspace{0.04\columnwidth}
\begin{minipage}[!ht]{0.48\columnwidth}
\centering
\includegraphics[trim=0cm 0cm 0cm 0cm, clip=true,angle=0,width=0.95\linewidth]{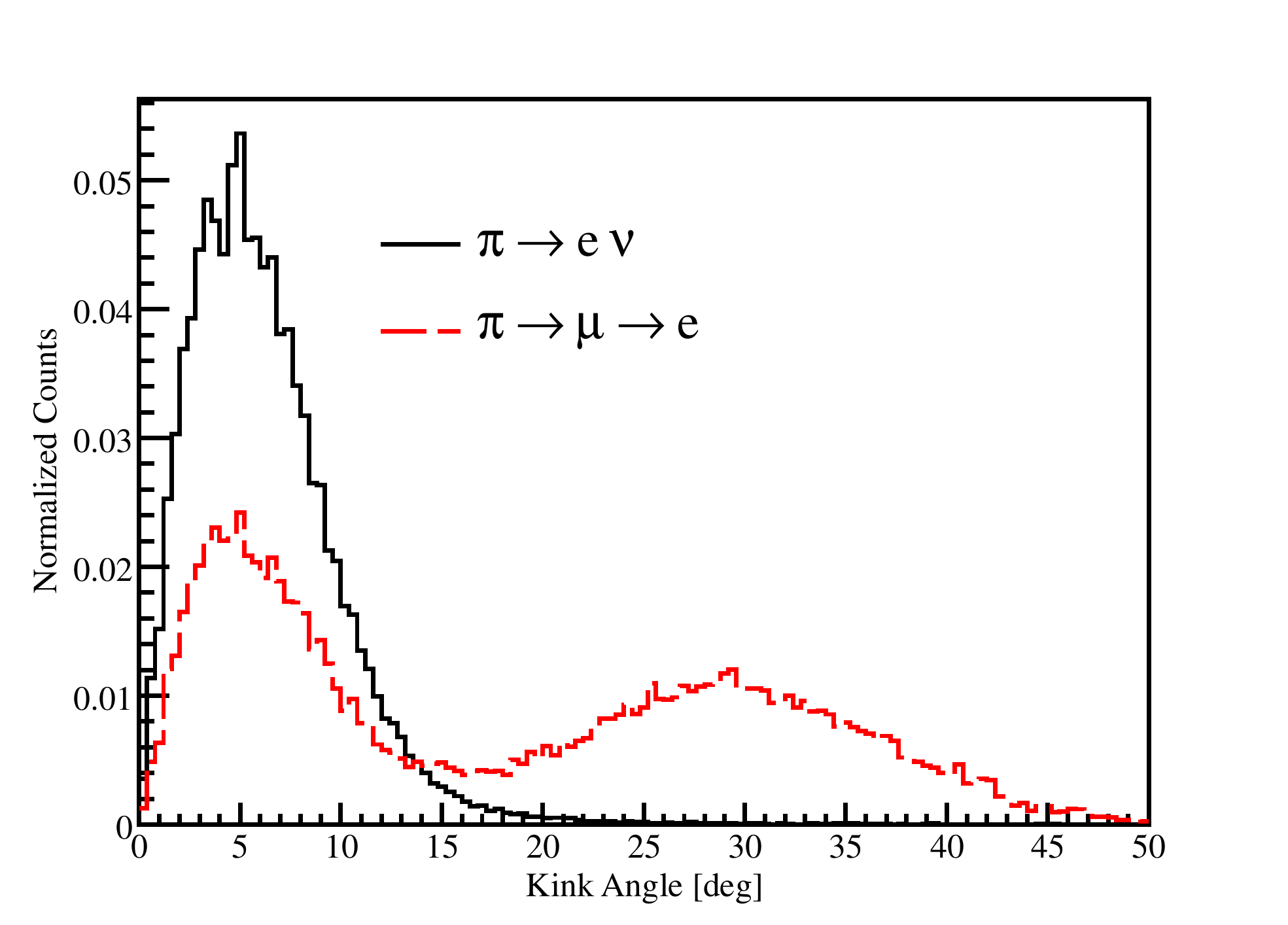}
\caption{Kink angle for \pienu (T$_{NaI(T\ell)}>55$ MeV) and $\pi^+\to\mu^+\to e^+$  (T$_{NaI(T\ell)}<30$ MeV) 
data events. The presence of pion-DIF events in the  $\pi^+\to\mu^+\to e^+$ sample can be clearly seen at angles $>$ 
15$^{\circ}$.} \label{fig:kinkdata}
\end{minipage}
\end{figure}

The distributions of the angle (kink angle)
between the track detected by WC1 and WC2 and the track detected by S1 and S2 for different decay types
are shown in Figure \ref{fig:kinkMC} for MC and on Figure \ref{fig:kinkdata} for data.
Pion-DIF events between WC2 and S1 can be distinguished due to their 
higher kink angle.

\section{Trigger and data acquisition system}
\subsection{\pienu run trigger}
Pions and a fraction of beam muons and positrons (used for the calibration of some detectors) 
were selected by the trigger using energy deposit information in B1. Definition of 
a beam particle was done by requiring the coincidence of the beam counters B1, B2, and B3. 
A coincidence of T1 and T2 counters defined the decay-positron signal. A coincidence of beam 
pion and decay positron signals 
(PIE) within a time window of --300 ns to 500 ns was
 the basis of the main trigger logic. Since \pimue ~decays happen much more often than \pienu 
decays, a {\it Prescale} trigger precisely selected only 1/16 of PIE events.
 Meanwhile, the \pienu events were enhanced by the {\it Early} and {\it HE} (High Energy) triggers. 
The {\it Early} trigger selected decays which happened between 2 ns and  40 ns 
(excluding prompt events) after the pion stop. Due to the 26 ns 
pion lifetime, more than 70\% of the \pienu events happen within this time range.
 The {\it HE} trigger was based on a VME-module which summed the energy 
deposited in the calorimeters in real time. The trigger signal was produced by this module
 for events which had a high energy deposited in the NaI(T$\ell$) and CsI spectrometers. The energy threshold was 
set 4~MeV below the highest energy of the positron from the $\pi^+\to\mu^+\to e^+$ decay
 chain.
  Almost all the \pienu events (with the exclusion of the tail events which extend below
 the $e^+$ energy spectrum from the $\pi^+\to\mu^+\to e^+$ decay chain) were selected by this trigger. Those three triggers constituted the
 ``physics triggers''. Additional triggers were used for data quality checks and calibration purposes.
The  {\it Xe-lamp} trigger 
provided flashes twice a second to all CsI crystals. Both these triggers ({\it Xe-lamp} and {\it Cosmic}) were intended 
for CsI calibration and monitoring. Finally,  the {\it Beam positron} trigger triggered 
on one of every 32 beam positrons to calibrate the NaI(T$\ell$) crystal. During a run, all 6 
triggers were used and several of them could be satisfied at the same time.
 The rates of the various triggers are shown in Table \ref{tb:TriggerRate}. 
The trigger signal issued by any of the six triggers was then latched by the pion 
(t$_{\pi^+}$) and the positron (t$_{e^+}$) timings in order to reference the gate of
 the data acquisition modules to the incoming and decay particles, respectively. These 
latched signals triggered the data acquisition. T$_{e^+}$ was used to trigger the data
 acquisition by the VME modules (VF48 and VT48, which are described below) while  T$_{\pi^+}$ triggered the COPPER
 data acquisition, described below. 

\begin{table}[htbp]
\begin{center}
\caption{Trigger rates}
\label{tb:TriggerRate}
\vspace{0.5cm}
\begin{tabular}{p{6cm}c}
\hline \hline 
Trigger&Rate [Hz] \\ \hline
\hline
Pion stop in target & 5$\times$10$^4$\\ \hline
\textit{Physics triggers}&\\
\hline
Early trigger&160\\
HE trigger&170 \\
Prescale trigger&240\\
\hline
\textit{Other triggers}&\\\hline
Cosmics trigger&15\\
Beam Positron trigger&5\\
Xe lamp trigger&2\\\hline
Total Triggers & 600\\
\hline
\hline
\end{tabular}
\end{center}
\end{table}

\subsection{Data acquisition system}
\label{sec:daq}
{\bf COPPER: }
This 500-MHz-sampling-frequency Flash ADC (FADC) system was based on the COPPER 
(COmmon Pipelined Platform for Electronics Readout) platform. The detailed characteristics
 of the COPPER system have been described in Refs.\cite{copper, IEEE2007}. The PIENU experiment
 was equipped with four COPPER boards\footnote{An additional COPPER board was installed during the 2012 run to record partial analog sums of the CsI crystal PMTs.} to digitize the signals coming from  the 23 PMTs reading 
out the plastic scintillators, analog sum of NaI(T$\ell$) PMTs, and partial analog sums of CsI
 array PMTs. The time window of the signals recorded by COPPER covered 7.74~$\mu$s 
(1.35~$\mu$s after and 6.39~$\mu$s before the pion stop) to be able to detect pre- and 
post-pile up particles. The waveforms recorded by COPPER were fitted, on an event-by-event
 basis, using templates obtained from a spline interpolation of the average PMT pulse shape.

\begin{figure}[!ht]
\begin{minipage}[!ht]{1\columnwidth}
\centering
\includegraphics[trim=0cm 0cm 0cm 0cm, clip=true,angle=90,width=\linewidth]{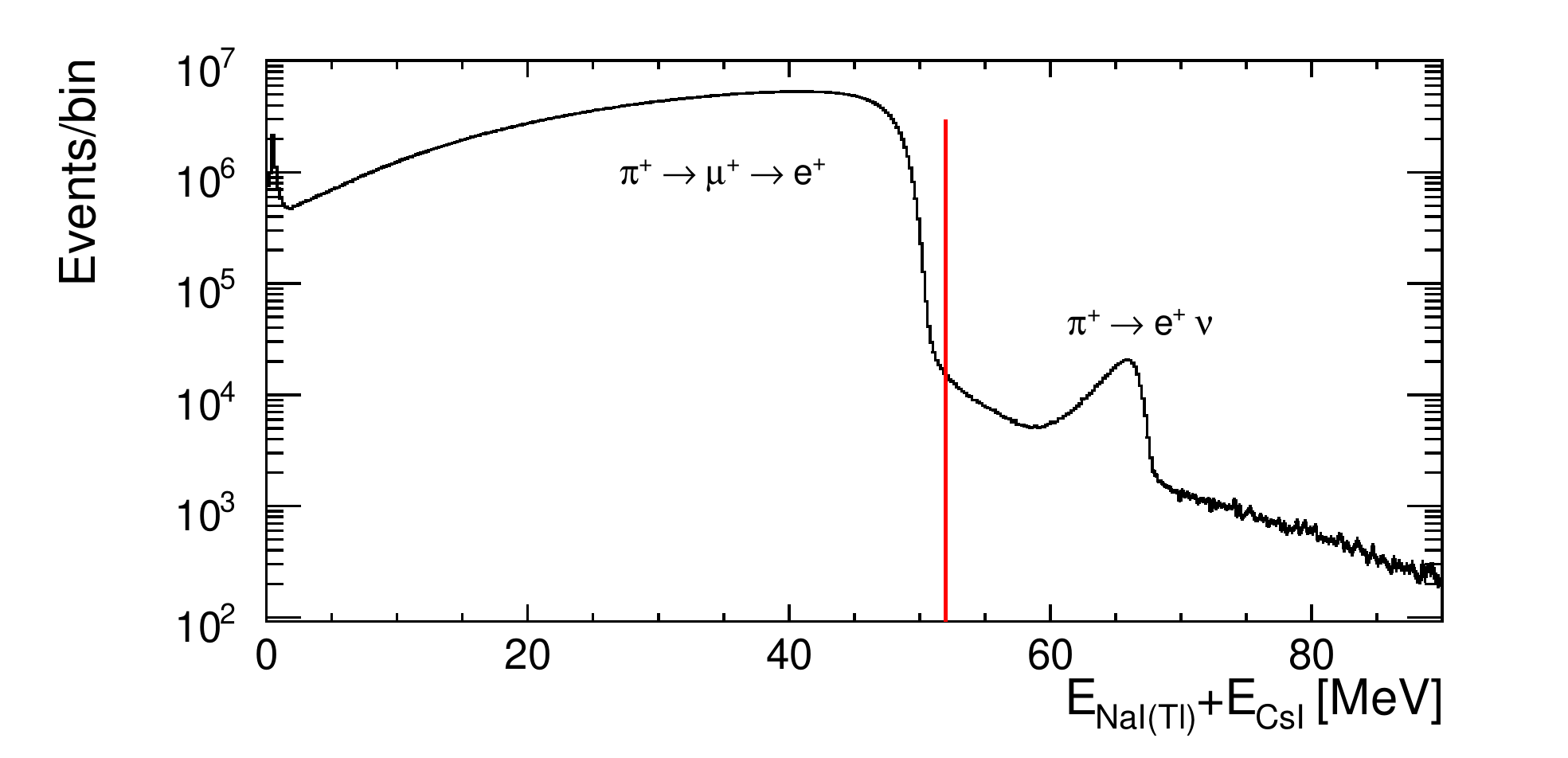}
\caption{Measured energy in NaI(T$\ell$) and CsI array for a portion 
of collected data. The vertical line
denotes separation of collected events into low-energy (dominated by \pimue $ $ decays) and high-energy (dominated by \pienu decays and pile-up).}
\label{fig:es_res}
\end{minipage}
\vspace{0.04\columnwidth}
\begin{minipage}[!ht]{0.48\columnwidth}
\centering
\includegraphics[trim=0cm 0cm 0cm 0cm, clip=true,angle=0,width=0.95\linewidth]{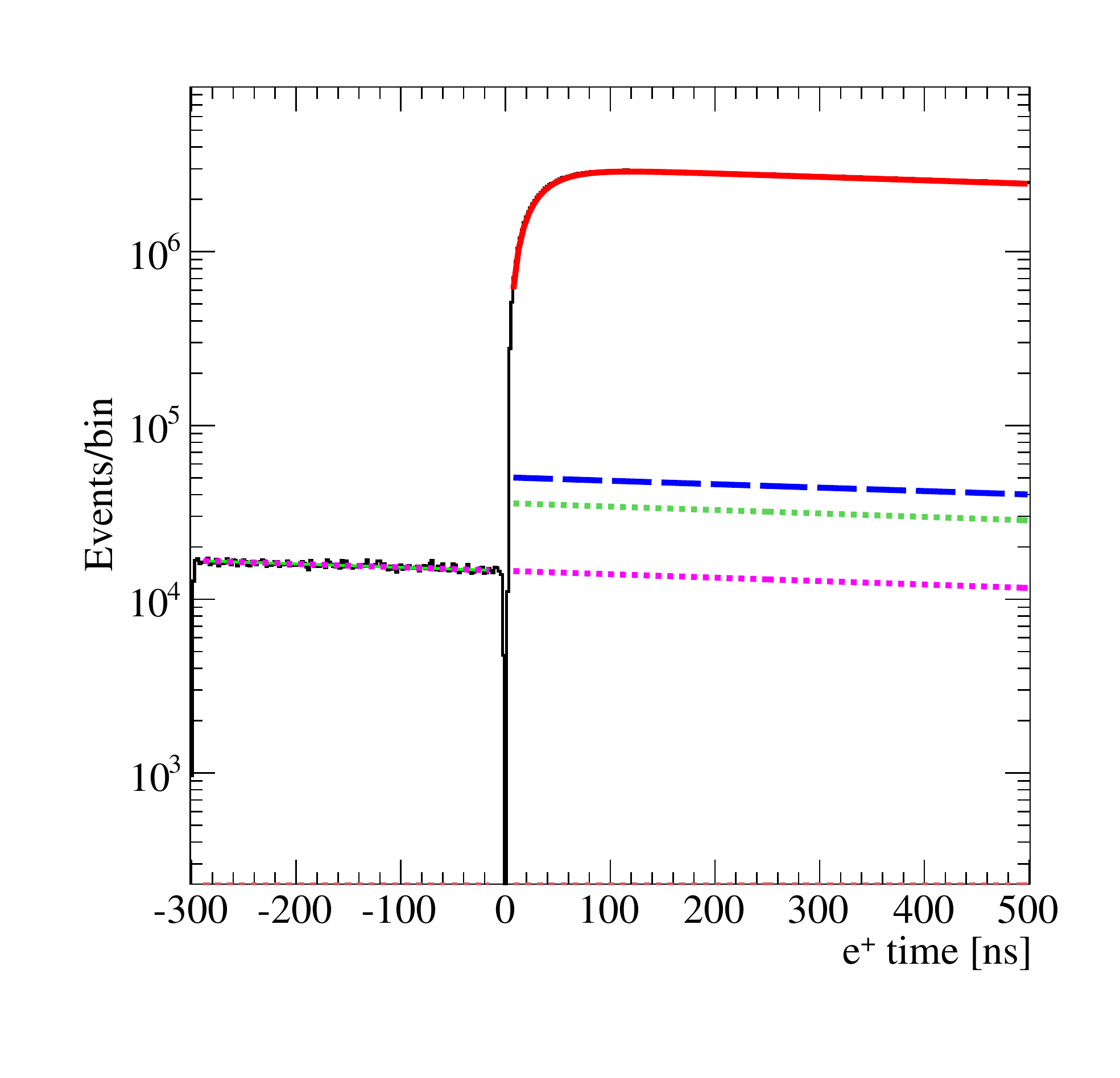}
\caption{Time spectrum defined by the time difference between T1 and B1 hits for low-energy events. Zero time
corresponds to the time of the pion stop. The solid (red) line represents the fit function describing the $\pi^+\to \mu^+ \to e^+$  decay chain, the dashed (blue) line represents the sum of backgrounds, represented by two dotted lines which correspond to muon decays and pion decays in flight.}
\label{fig:ts_le}
\end{minipage}
\hspace{0.04\columnwidth}
\begin{minipage}[!ht]{0.48\columnwidth}
\centering
\includegraphics[trim=0cm 0cm 0cm 0cm, clip=true,angle=0,width=0.95\linewidth]{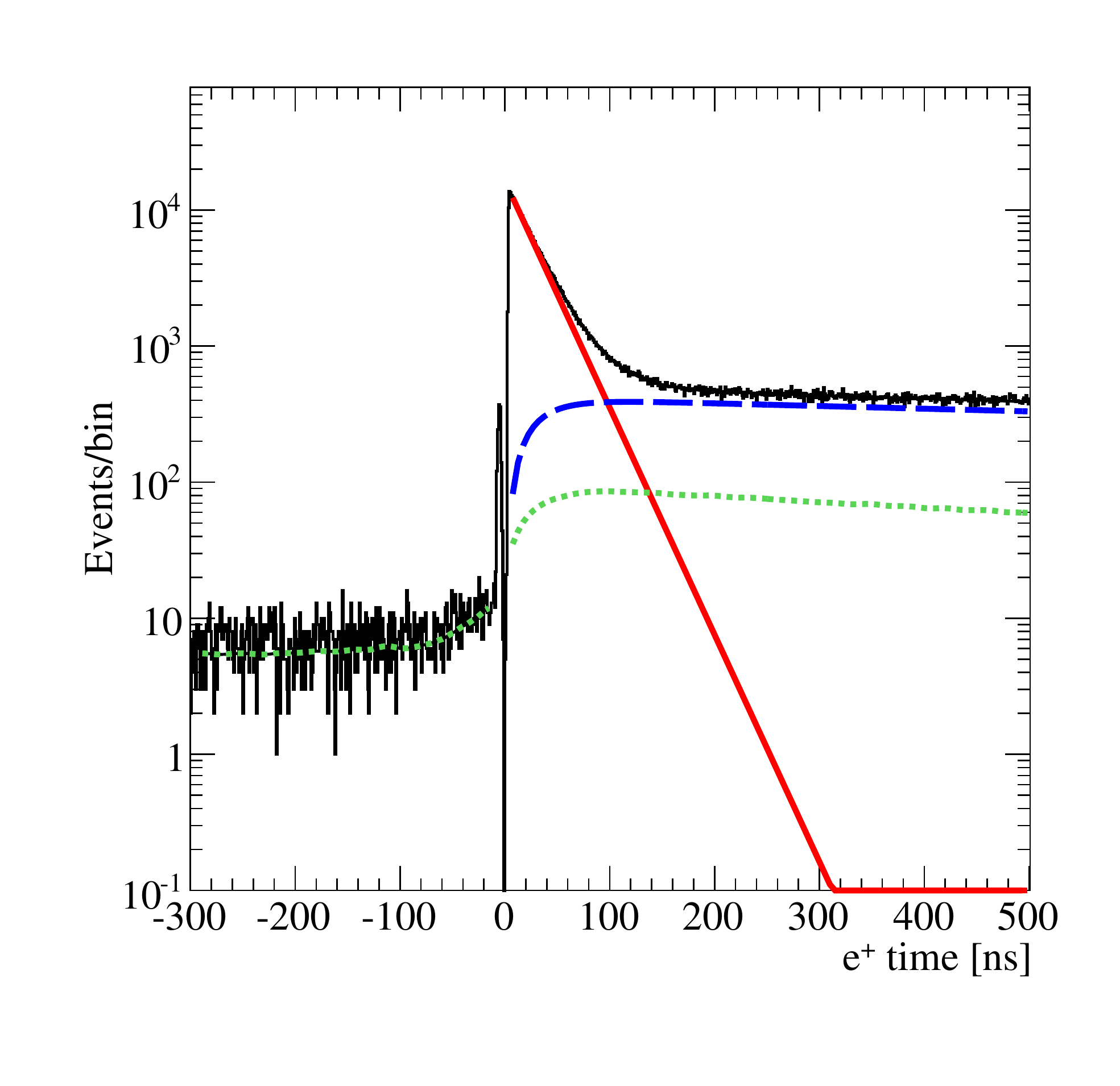}
\caption{Time spectrum defined by the time difference between  T1  and B1  hits for high-energy events. Zero time corresponds to the time of the pion stop. The solid (red) line represents the fit function describing the \pienu time spectrum, the dashed (blue) line represents  the $\pi^+\to \mu^+ \to e^+$ decay chain events which due to pulse pile up ended up in the high energy spectrum, and the dashed line represents the sum of the remaining background mechanisms like charged pileup, radiative decays, etc.}
\label{fig:ts_he}
\end{minipage}
\end{figure}
{\bf VF48: }
The VF48 is a 48 channel 60 MHz FADC \cite{vf48}. 
It has a resolution of
 10 bits and a dynamic range of $\pm$250 mV. All the NaI(T$\ell$) and the CsI PMT signals as well
 as all the silicon strip channels were read out by VF48 modules. This made a total of 404 channels 
(NaI(T$\ell$): 19, CsI: 97, silicon strips: 288) read out by 10 VF48 modules). Typically, the VF48 read out 40 samples around the time defined
 by the decay positron. All VF48 modules were fed with the same 20-MHz clock provided by
 another VME module. This clock was multiplied internally in the VF48 to reach 60-MHz
 sampling. Data suppression was implemented in all channels except those reading the
 NaI(T$\ell$) crystal PMTs. Additionally, due to the slow NaI(T$\ell$) response, its waveforms 
were digitized at a rate of 30~MHz instead of 60~MHz. 

{\bf VT48: }
Discriminated signals from the three wire chambers (WC1, WC2, and WC3), all the logic signals issued by the PMT 
signals after discrimination, and most trigger logic signals were read out by multi-hit 
time-to-digital converter 
\cite{vt48}.
 This device, based on the AMT-2 chips \cite{AMT2}, has a 25-MHz on-board clock which is multiplied to achieve 0.625~ns resolution.
 All VT48s, however, were fed with an external 25-MHz clock to ensure the synchronization
 of modules. One board can read out 48 channels for up to 20~$\mu$s. To optimize the
 dead-time, only two channels were read out for 20~$\mu$s to detect long-lifetime backgrounds
 originating from the beam while all other channels were read out 4.0~$\mu$s before
 and after the time defined by the decay positron signal in T1. In total, 11 VT48 modules
 were used in the experiment.

The PIENU data acquisition system consisted of three VME crates (two VMEs hosted the VF48 and
 VT48 modules while the third mostly ran slow control modules and modules used by the
 COPPER system) controlled by VME master modules, and four COPPER boards with a processor on
 each board.  Each processor was running the associated front end programs to transfer
 the data to a host computer.
Collection of the data was done by the MIDAS data acquisition system which incorporates
 an integrated slow control system with a fast online database and a history system
 \cite{MIDAS}. The MIDAS server computer can be controlled via a web interface. All the
 information and errors from the DAQ modules were issued on the web page and programs
 checking the quality of the data online were connected to MIDAS during data taking.

\section{Conclusion}

The PIENU detector was commissioned in 2009 and was operated during physics and
 supplemental measurements including NaI(T$\ell$) response measurements until the end of 2012. 
The branching ratio $R^{e/\mu}$ will be determined from the observed numbers of $\mathrm\pi^+\rightarrow\mathrm e^+{\mathrm\nu}_e$ and $\mathrm\pi^+\rightarrow\mathrm\mu^+\rightarrow\mathrm e^+$ events. The summed energy in the NaI(Tℓ) and CsI array for events in the time window -300 to +500 ns in relation to the pion stop time is shown in Fig. \ref{fig:es_res} where the energy spectrum is divided into two regions at 52 MeV:  low-energy events dominated by  $\mathrm\pi^+\rightarrow\mathrm\mu^+\rightarrow\mathrm e^+$ decays,  and high-energy events dominated by  $\mathrm\pi^+\rightarrow\mathrm e^+{\mathrm\nu}_e$ decays. The time spectra for these two regions shown in Figs. \ref{fig:ts_le} and \ref{fig:ts_he} are used to determine the raw branching ratio taking into account various background components such aspile-up hits from previous muon decays, and pion decays-in-flight. The $R^{e/\mu}$ is then determined after various corrections are made (See Section 1). 
The excellent performance of  the crystal array 
and
 tracking array, paired with good time
 resolution  achieved
 in plastic scintillators (B1, T1) were important aspects of the PIENU apparatus
 ensuring the possibility of a successful measurement of \BRs ~and improved 
searches for heavy neutrinos \cite{mnu}.

\section{Acknowledgment}
We wish to thank P. Amaudruz, R. Bula, S. Chan, M. Constable, C. Lim, N. Khan, R. Kokke, P. Lu, K. Olchanski, R. Openshaw, C. Pearson and R. Poutissou
 for their contributions to the engineering, 
installation and commissioning of the detector and DAQ-related work. We are also grateful to
Brookhaven National Laboratory for providing the NaI(T$\ell$) and
CsI crystals.  This work was supported by the Natural Science and
Engineering Council (NSERC) and the National Research Council of
Canada through its contribution to TRIUMF and supported by JSPS KAKENHI Grant Number 18540274, 21340059, 24224006.



\bibliographystyle{elsarticle-num}
\bibliography{pienu}


\end{document}